\documentclass[10pt,journal,a4paper,twoside]{./IMEJ}

\usepackage{cite}
\usepackage[pdftex]{graphicx}
\usepackage{textcomp}
\usepackage{amsmath}
\usepackage{booktabs}
\usepackage{threeparttable}
\usepackage[caption=false,font=footnotesize]{subfig}
\usepackage{url}
\usepackage{balance}
\usepackage[english]{babel}
\usepackage{caption}
\bstctlcite{IEEEexample:BSTcontrol}

\usepackage{amsmath}
\usepackage{amssymb}
\usepackage{graphicx}
\usepackage{multicol}
 \usepackage{multirow}
\usepackage[colorlinks=true, allcolors=blue]{hyperref}

\title{Benchmarking machine learning models for predicting aerofoil performance}

\author{Oliver Summerell, Gerardo Aragon-Camarasa and~Stephanie Ordonez Sanchez
\thanks{\textcopyright~2025 European Wave and Tidal Energy Conference. This paper has been subjected to single-blind peer review.}%
\thanks{This work was supported in part by the EPSRC under grant 2890149}
\thanks{O. Summerell and G. Aragon-Camarasa are with the CVAS Group in the Computer Science Dept. of the University of Glasgow, University Avenue, Glasgow, G12 8QQ (email: o.summerell.1@research.gla.ac.uk)}
\thanks{S. Ordonez Sanchez is with the Mechanical and Aerospace Engineering Dept. of Strathclyde University, 16 Richmond Street, Glasgow, G1 1XQ (email: s.ordonez@strath.ac.uk)}%
\thanks{Digital Object Identifier:\\\protect\url{https://doi.org/10.36688/ewtec-2025-879} }
} 

\begin{document}
\renewcommand{\thepage}{879--\arabic{page}}
\maketitle

\begin{abstract}

This paper investigates the capability of Neural Networks (NNs) as alternatives to the traditional methods to analyse the performance of aerofoils used in the wind and tidal energy industry. The current methods used to assess the characteristic lift and drag coefficients include Computational Fluid Dynamics (CFD), thin aerofoil  and panel methods, all face trade-offs between computational speed and the accuracy of the results and as such NNs have been investigated as an alternative with the aim that it would perform both quickly and accurately. As such, this paper provides a benchmark for the \texttt{windAI\_bench} dataset published by the National Renewable Energy Laboratory (NREL) in the USA.
In order to validate the methodology of the benchmarking, the \texttt{AirfRANS} dataset benchmark is used as both a starting point and a point of comparison.

This study evaluates four neural networks (MLP, PointNet, GraphSAGE, GUNet) trained on a range of aerofoils at 25 angles of attack (4$^\circ$ to 20$^\circ$) to predict fluid flow and calculate lift coefficients ($C_L$) via the panel method.
GraphSAGE and GUNet performed well during the training phase, but underperformed during testing.
Accordingly, this paper has identified PointNet and MLP as the two strongest models tested, however whilst the results from MLP are more commonly correct for predicting the behaviour of the fluid, the results from PointNet provide the more accurate results for calculating $C_L$.


\end{abstract}

\section{Introduction}
\label{intro}

In an age of rising energy demands, tidal stream technologies have the potential to supply a significant level of clean and renewable energy, with the UK alone having an estimated, practical resource of 34TWh/year~\cite{Coles2021}. Whilst this is equivalent to around 11\% of the national energy consumption of the UK, very little of it has been realised at the point of writing \cite{Coles2021}. 

One of the key reasons for the lack of tidal energy being tapped in the UK is the cost attributed to the design and installation of the turbines. The \texttt{Contracts for Difference} scheme, a UK government scheme for the support of low-carbon energy production, priced tidal stream energy at £172/MWh in 2024, compared to offshore wind, which costs £58.87/MWh or £139.93/MWh for floating offshore~\cite{CfD24}. 

The process of designing a tidal turbine requires a variety of methods to evaluate and optimise their operation. The most commonly used methods include a variety of versions of Computational Fluid Dynamics (CFD). Faster and less computationally expensive methods include the use of Blade Element Momentum theory (BEMT)~\cite{MANNION2020106918}. BEMT, however, relies on a number of correction factors and databases to account for the aero/hydrodynamics of the aerofoil distribution along the blade~\cite{MAHMUDDIN20171123}. \texttt{XFoil} is often used to obtain values, as it provides a balance between computational speed and accuracy. As shown in Morgado et al. \cite{MORGADO2016207}, \texttt{XFoil} runs quickly and accurately, but it experiences a dramatic drop in accuracy when subjected to high $Re$ or high angles of attack compared to CFD methods. Furthermore, \texttt{XFoil} is unable to `account for transition with Navier-Stokes solvers'~\cite{MORGADO2016207}, reducing its usefulness for more complex use cases.

As a result, the last few years have seen Neural Networks (NNs) being proposed as a potential alternative to the current methods, with the aim to provide comparable results to CFD models whilst requiring lower computational cost~\cite{cfdbench}. One of the main drawbacks of using NNs is the reliance on large quantities of reliable data to be trained on, and due to the computational intensity of CFD simulations, few relevant datasets are available. That is to say, whilst there are a variety of fluid flow datasets available, such as \texttt{CFDBench}~\cite{cfdbench} and \texttt{MegaFlow2D}~\cite{megaflow2d}, they do not contain data for aerofoils and, therefore cannot be used for the purposes of aerofoil performance prediction.

There are two datasets being considered within this paper, firstly is \texttt{windAI\_bench} \cite{windAi_bench}, published by NREL, which contains a large number of aerofoils, more than sufficient for use in NNs. The actual layout of this data is explored in section \ref{meth:data}.
To the knowledge of the authors, no public benchmarking has been performed using the \texttt{windAI\_bench} dataset with contemporary models.
As such, this paper aims to provide a benchmark for \texttt{windAI\_bench} to verify its applicability within the field.

In order to validate the methodology applied to benchmarking \texttt{windAI\_bench}, the second dataset being investigated is the \texttt{AirfRANS} dataset published by Bonnet et al.~\cite{AirfRANS}. \texttt{AirfRANS} contains 1,000 aerofoils and is benchmarked for use in Machine Learning (ML) by using four separate architectures. 
Importantly, whilst it includes training and validation loops, \texttt{AirfRANS} does not incorporate testing against unseen geometries or foils, meaning that the ability of any of the benchmark models to generalise to completely unseen data has not been verified.
With this in mind, whilst \texttt{AirfRANS} will be of great use for a comparison of methodologies used for benchmarking this type of data, it can not be used for comparison of results due to the fundamental difference between the two.
To keep the results in line with those for similar datasets, the models used will be the same as those in Bonnet et al.~\cite{AirfRANS}.

One point worth noting is that whilst these datasets are designed for wind turbines, they are also applicable to the design and evaluation of tidal turbines. To account for scalability issues, Reynolds number calculations would thus need to consider flow characteristics and operational velocities at which tidal turbines usually operate.

As mentioned, the computational requirement of generating CFD data means that the data is regularly unattainable in large enough quantities for data-driven NNs to be viable. This paper aims to investigate the amount of data required to attain reasonable results. Therefore, the performance of each model has been investigated over 5, 20, 55 and 150 aerofoils. These numbers were chosen to be roughly equal to an exponential scale as $5 \approx e^{1.5}$, $20 \approx e^3$, $55 \approx e^4$ and $150 \approx e^5$. It is important to validate the results from this against a sample set of aerofoils not used in the training of the model, i.e., a test dataset, to ensure that the model is not over-fitting to the data presented.

\section{Related Literature}
\label{LitRev}

\begin{figure}
    \centering
    \includegraphics[width=1\linewidth]{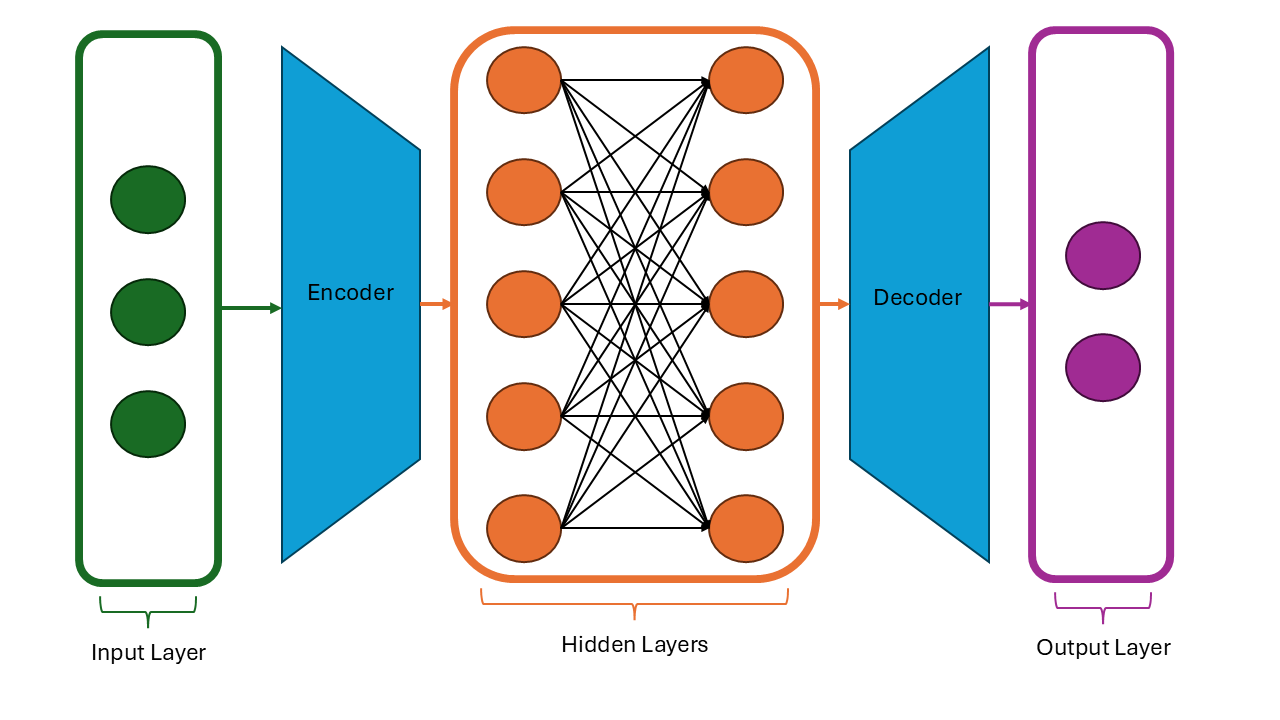}
    \caption{A Fully Connected Neural Network with an encoder and decoder}
    \label{fig:NN}
\end{figure}

Artificial Neural Networks (ANNs), or simply NNs, are universal function approximators~\cite{HORNIK1989359} used within ML for the prediction of specified functions. NNs are called so because they are based on the design of the biological neural networks found within a brain, with neurons and synapses represented by the neurons that make up each layer of the NN and the connections that connect them, a diagram of which can be seen in Fig. \ref{fig:NN}. The network is divided up into three categories of layers (input, hidden and output) each of a user specified width, i.e. how many neurons in each layer. Each connection contains a weight ($W$) which can be within the range $0 \leq W \leq 1$. These are multiplied by the output from the neuron it comes from ($x$) before being summed up with the remaining connections along with a bias factor ($b$), as seen in (\ref{eq:neuron}).

\begin{equation}
    y = b + \sum_{i=1}^n W_ix_i
    \label{eq:neuron}
\end{equation}
where $y$ is the value of the new node, $n$ is the number of nodes in the previous layer.

These weights and neuron values are used to affect the final output of the NN and this output is then compared to a `ground truth' value, i.e. the `real' value usually taken from the data. Then, after finding the gradient of the `loss', or difference between these two values, the weights and biases of the network are updated using an optimisation algorithm such as ADAM~\cite{adam}.
The data loaded into the model is split into training and validation data: the former is used to optimise the weights and biases of the model: whilst the latter is used to check the ability of the model to generalise to unseen data. It is worth noting that incorporating a validation split of data is not sufficient to claim the model can generalise to any unseen data. It is used to ensure that the model does not `over-fit' to the training data.
This is then repeated until a point defined by the user, usually either when a certain number of epochs, or iterations, have been performed or when the loss has gone below a pre-determined threshold.

Some NNs make use of an encoder and a decoder within the code. The encoder block takes the input data and abstracts it into a \textit{latent} vector, similar to the first two layers in Fig. \ref{fig:NN}, whereas the decoder takes the latent vector and aims to reconstruct the input as the output. In this paper, there is an encoder before each model and a decoder after it. They are both defined using MLP and are trained along with the model.

There are a variety of ways in which NNs have been implemented in the disciplines surrounding renewable energies and fluid flow. In \cite{Wang2022}, NNs are used to model the flow of fluid within a combustor. The model uses experimental data from Particle Image Velocimetry (PIV) in order to predict the behaviour of said fluid. An older paper, \cite{Kalogirou2001}, published in 2001, looks at a variety of applications for NNs within the renewable energy industry, but with a focus on solar power. The model performed notably well across all experiments, with a loss of between 1-9\% accuracy.

One variation on the typical format of the NN is the Graph Neural Network (GNN)~\cite{OGGNN}, which differs slightly from a regular NN as the data is defined with nodes connected to each other via edges, similarly to the layout of a mesh. As such, to determine the values of each node, the information is propagated to and from the neighbouring nodes iteratively.~\cite{gnn_survey}

GNNs have been increasingly used for the prediction of fluid dynamics such as in Li et al.~\cite{Li2023}, and Peng et al.~\cite{Peng22_gridGCNN}.
Ogoke et al.~\cite{Ogoke2021} use Graph Convolutional Networks (GCNs), first introduced by Kipf et al.~\cite{Kipf2016}, to predict the drag coefficient ($C_D$), from a similar input to both \texttt{windAI\_bench} and \texttt{AirfRANS}. The main difference in methodology that Ogoke et al. use is they are predicting solely $C_D$, and not the entire fluid behaviour.

The dataset being used to compare the benchmarks of \texttt{windAI\_bench} \cite{windAi_bench} to is \texttt{AirfRANS}, published by Bonnet et al.~\cite{AirfRANS}, due to its pre-existing benchmark of fluid flow over an aerofoil. It includes 1,000 aerofoils that have been benchmarked using four architectures: a basic MLP, PointNet \cite{pointnet}, GraphSAGE \cite{hamilton2018}, and GUNet \cite{GUNet}, which is a GNN-based variant of the UNet architecture~\cite{Ronneberger2015}. As mentioned in the introduction, further insight is given into the two datasets in section \ref{meth:data}.

\section{Methodology}
\label{Meth}
The code for this paper was done using \texttt{PyTorch} and \texttt{PyTorch Geometric} \cite{Fey2019} and is available on GitHub: \url{https://github.com/OllieS-PhD/Benchmark_Aerofoils}
\subsection{Data}
\label{meth:data}

\texttt{WindAI\_bench} includes two different models, transition and turbulent, both run by a 2D Hamiltonian Strand Navier-Stokes solver (HAM2D). Specifically being used is the \texttt{airfoil\_2k} subsection of the dataset, which contains 1830 unique aerofoil shapes at 25 angles of attack, ranging from $-4^\circ$ to $20^\circ$, for 3 separate Reynolds numbers. This gives a total of $45,750$ simulations to train with, making the dataset comparable to that used by \cite{Bouhlel2020}. In order to give an insight into the amount of data required for reasonable results to be achieved by the models, only a limited number of aerofoils (across every angle of attack (AoA)) were utilised within this paper.
Four sets of simulations was performed with 5, 20, 55 and 150 aerofoils for each of the models, the number of foils for each set has been chosen in line with an exponential scale, such that: $5 \approx e^{1.5}$, $20 \approx e^3$, $55 \approx e^4$ and $150 \approx e^5$. Whilst a full breakdown of the data can be found on \texttt{windAI\_bench}'s GitHub \cite{windAi_bench}, for the purposes of this paper, the data used was the $Re = 3e^6$ data generated by the turbulence model. This includes, for each AoA, the $C_L$ and $C_D$ values, the `landmarks' array which is a coordinate file for each point along the aerofoil, and the physical parameters at each node: namely the $x$ and $y$ positional co-ordinates, the density of the fluid, the $x$ and $y$ components of the momentum, the total energy, and the vorticity.

The \texttt{airfoil\_2k} data was stored as a single H5 data structure, from which the data relevant to this paper was processed and stored into separate H5 files, one for each aerofoil.
H5 files make use of HDF5, a `Hierarchical Data Format' that is able to store large amounts of data that can be easily read and indexed~\cite{h5}.

The dataset was also normalised, making it non-dimensional as this reduces any potential inaccuracies caused by performing mathematical operations on largely differing values. The normalised values are found as in (\ref{eq:eq1}).

\begin{equation}
\begin{array}{ccc}
\displaystyle \rho^* = \frac{\rho}{\rho_{\infty}}; &
\displaystyle \sigma^* = \rho^*\frac{U}{a_{\infty}}; & 
\displaystyle \omega^* = \frac{\omega}{a_{\infty}}
\end{array}
\label{eq:eq1}
\end{equation}

where, $\rho$ is density, $\sigma$, momentum, $U$, velocity magnitude, $\omega$, vorticity, $a$, speed of sound, and $\infty$ denotes the free-stream conditions, which in this case means $\rho_\infty = 1.225kgm^{-3}$ and $a_\infty = 340.15ms^{-1}$. As such, the momentum is given with respect to its relative Mach Number ($Ma$), meaning the magnitude of the free-stream momentum is equal to $0.1 Ma$ within the simulations used for training.

As the interactions between the fluid and the foil determine the aerodynamic properties of the aerofoil, the most important points in the domain are those at the surface of the aerofoil itself. The geometry information was not present within the mesh itself, but was defined as an array of 2D Cartesian co-ordinates in a separate geometry file labelled `landmarks'. As such, determining which points lay on the surface was done by using the Python package, \texttt{Shapely.Polygon}.~\cite{shapely}
A polygon object is generated using the `landmarks' data, resulting in a shape made of straight lines going from point to point instead of the smooth curve expected of a foil. As such, none of the mesh points lie on the polygon itself and a buffer zone is implemented in order to check which points are closest to the foil. Whilst this inclusion of the buffer zone works with a high accuracy, it is not without drawbacks. Most often happening at the leading edge of the aerofoils, there are often a couple of extra nodes `caught' in the buffer zone. However, after testing a range of buffer distances, $3.5e^{-6}m$ was found to be the best compromise between losing nodes at the upper and lower surfaces and catching extra nodes at the leading edge.
Introducing this separation between surface, volume, and total areas was also implemented in the \texttt{AirfRANS} paper, and as such, it allows for a comparison of the effect of training on these areas and further comparison of the two benchmarks.

\begin{figure}
    \centering
    \includegraphics[width=1\linewidth]{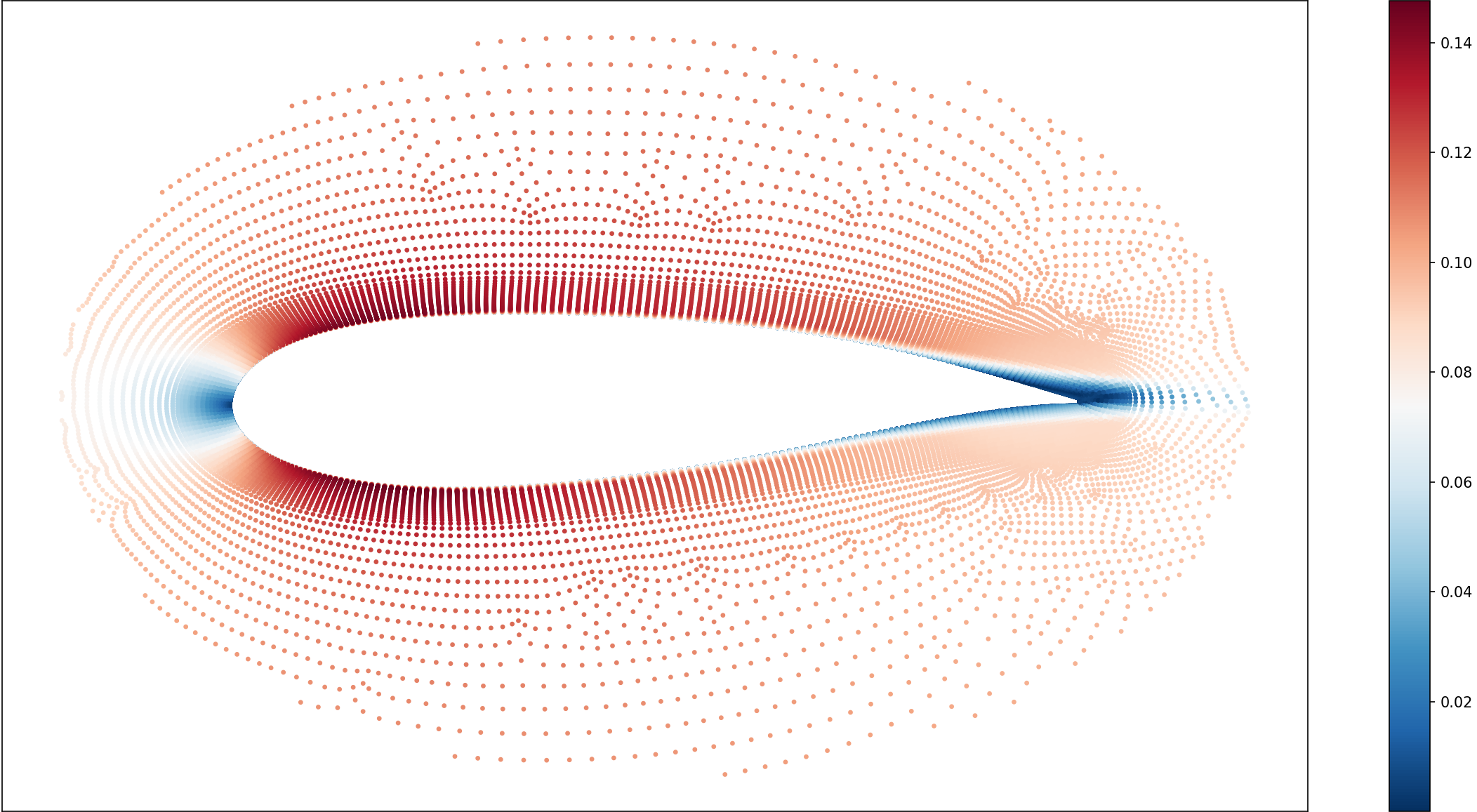}
    \caption{Point cloud of aerofoil \#18 with reduced nodes for training, showing the ground truth for the Magnitude Momentum at an AoA of 0\textdegree}
    \label{fig:red_view}
\end{figure}

The errors on the nodes farther away from the aerofoil were typically very low when compared to those close to the aerofoil, causing the overall error of the model to be artificially low, creating a large error at the points in which this paper is interested. Therefore, when loading the data, all points outside a radius of $0.7 m$, centred at the midpoint of the chord length ($c=1 m$), were omitted, noticeably increasing the error of the system, but ensuring better results at the points of interest. A graphical interpretation of this can be seen in Fig. \ref{fig:red_view}.

\subsection{Models}
\label{meth:mod}

In order to ensure that the results are comparable to those presented in other datasets, models have been taken from the \texttt{AirfRANS} dataset benchmark paper \cite{AirfRANS}.
One of the most notable differences between the two datasets is the properties used to define the fluid flow. \texttt{AirfRANS} has the models outputs/targets as velocity, pressure, and viscosity at each node, while \texttt{windAI\_bench} uses density, momentum, energy, and vorticity.
The inputs for this model are then based on the requirements of the outputs and for each node containing: positional co-ordinates, free-stream density, inflow momentum components, the initial energy and vorticity of the system set to $\varepsilon$ ($=1e^{-10}$), and the negative signed distance function which determines the distance from the node to the aerofoil.~\cite{SDF} The data was then normalised again using just the mean and standard deviation of the training dataset to reduce errors in the training.

The loss criterion used for the models during training is the Mean Squared Error (MSE) as outlined in (\ref{eq:mse}).

\begin{equation}
\displaystyle MSE=\frac{1}{n}\sum_{i=1}^n (Y_i-\hat{Y}_i)^2
\label{eq:mse}
\end{equation}

where, $Y$ is the ground truth value from the dataset and $\hat{Y}$ is the predicted value.

\subsection{Post Processing}
\label{meth:postproc}

To properly assess the effectiveness of the trained models, testing is performed using 60 unseen aerofoils to assess the ability of the models to generalise to relevant unseen data. The metrics of interest for this analysis are the root mean squared error (RMSE) of both the prediction and the $C_L$ calculated from the predicted flow. RMSE is more favourable than MSE as it converts the results back to the original scale and units of the target value, making it an easier metric to interpret. The formulation for the RMSE can be seen in (\ref{eq:rmse}).

\begin{equation}
\displaystyle RMSE=\sqrt{MSE}=\sqrt{\frac{1}{n}\sum_{i=1}^n (Y_i-\hat{Y}_i)^2}
\label{eq:rmse}
\end{equation}

Using the predicted values of vorticity at the aerofoil, $C_L$ is calculated using the panel method~\cite{Liburdy2021} and then compared to the values provided by the \texttt{windAI\_bench} dataset via the RMSE.

The panel method is a method that allows for the approximation of forces acting upon an object in a flow by segmenting that object, in this case an aerofoil, into discrete surface elements, panels, and prescribing a flow element to it~\cite{Liburdy2021}. Using the volumetric vorticity ($\omega$) predicted at each node on the aerofoil, the total circulation ($\Gamma$) for the aerofoil can be calculated by integrating $\omega$ over the total area defined by the aerofoil as seen in (\ref{eq:Gamma}).

\begin{equation}
\displaystyle \Gamma = \iint_{A} \omega .dA
\label{eq:Gamma}
\end{equation}

This allows for the calculation of the Lift ($L$) and coefficient of lift $C_L$ acting on the aerofoil to be calculated as seen in \ref{eq:lift} \& \ref{eq:CL}~\cite{Liburdy2021}: 

\begin{equation}
\displaystyle L = \rho U_{\infty} \Gamma
\label{eq:lift}
\end{equation}\begin{equation}
\displaystyle C_L = \frac{L}{\frac{1}{2}\rho U_{\infty}^2 c} = \frac{2 \Gamma}{U_{\infty}c}
\label{eq:CL}
\end{equation}

where $U_{\infty}$ is the free stream velocity, $\rho$ is the fluid density and $c$ is the chord length, which in this case is $1$.

\section{Results}
\label{results}

As mentioned earlier in Section \ref{meth:mod}, MSE was used to calculate the loss during the training cycle, as is common practice, after which the final losses were converted into RMSE and presented in the tables below, in order to provide a more insightful view into accuracy.
The tables below are laid out as follows: for each total number of aerofoils, the first table contains the RMSE during training at points on the aerofoil surface (Surface) and across all other points that are not touching the aerofoil (Fluid). Then, the last three columns of the first table contain the results from testing as laid out in Section \ref{meth:postproc}, with the RMSE at the surface and in the fluid (the same points highlighted in the test RMSE) as well as the RMSE of the prediction of $C_L$ averaged over all 60 aerofoils (1,500 simulations).
The second and third sections for each table contain a breakdown of the RMSE during testing of each of the variables being predicted, namely density ($\rho$), x-momentum ($\sigma_u$), y-momentum ($\sigma_v$), energy ($e$), and vorticity ($\omega$).

Table \ref{tab:5f} shows the results for the tests trained on 5 aerofoils, Table \ref{tab:20f} for those trained on 20, Table \ref{tab:55f} for those trained on 55, and Table \ref{tab:150f} for those trained on 150.

It is worth noting that each number of aerofoils as displayed is not the same as the amount of data points the models are trained on as each foil has data for 25 AoAs. Therefore, the number of simulations each is trained on is: 125 for Table \ref{tab:5f}, 500 for Table \ref{tab:20f}, 1375 for Table \ref{tab:55f}, 3750 for Table \ref{tab:150f}.

For all results, a lower score is better, and the lowest for each section has been highlighted in bold. All simulations were run on Windows 11 Home with PyTorch version 2.5.1, utilising Intel(R) Core(TM) i9-10980XE CPU @ 3.00GHz, NVIDIA GeForce RTX 3080 GPU, and 120GB of RAM.

\begin{table}
\caption{Table of RMSEs produced from models training and testing on 5 aerofoils}
{
\centering
\resizebox{\columnwidth}{!}{\begin{tabular}{ |c||c c|c c c|  }
 \hline
 \multicolumn{6}{|c|}{\textbf{5 Aerofoils}} \\
 \hline
Model & \multicolumn{2}{c|}{Train RMSE}  & \multicolumn{3}{c|}{Test RMSE} \\
& Surface & Fluid & Surface & Fluid & CL \\
 \hline
 \textbf{MLP}          & 1.135 & 0.470 &  \textbf{1.630} & 0.510 & \textbf{0.873}  \\
 \textbf{PointNet}     & 1.298 & 0.448 &  1.703 & \textbf{0.483} & 1.019  \\
 \textbf{GraphSAGE}    & 0.718 & 0.342 &  2.157 & 1.109 & 1.020  \\
 \textbf{GUNet}        & \textbf{0.707} & \textbf{0.337} &  2.340 & 0.812 & 1.568  \\
 \hline
\end{tabular}}\par
}
{
\centering
\resizebox{\columnwidth}{!}{\begin{tabular}{ |c||c c c c c|  }
 \hline
 \multicolumn{6}{|c|}{\textbf{Test RMSE Per Variable at Surface}} \\
 \hline
 Model & $\rho$&$\sigma_u$&$\sigma_v$&$e$&$\omega$ \\
 \hline
 \textbf{MLP}           & 0.716 & \textbf{0.785} & \textbf{0.816} & 0.812 & \textbf{3.292} \\
 \textbf{PointNet}      & \textbf{0.690} & 0.914 & \textbf{0.816} & \textbf{0.784} & 3.452 \\
 \textbf{GraphSAGE}     & 1.210 & 0.802 & 0.868 & 1.406 & 4.294 \\
 \textbf{GUNet}         & 1.313 & 1.221 & 0.641 & 1.522 & 4.629 \\

 \hline
\end{tabular}}\par
}
{
\centering
\resizebox{\columnwidth}{!}{\begin{tabular}{ |c||c c c c c|  }
 \hline
 \multicolumn{6}{|c|}{\textbf{Test RMSE Per Variable in Fluid}} \\
 \hline
 Model & $\rho$&$\sigma_u$&$\sigma_v$&$e$&$\omega$ \\
 \hline
 \textbf{MLP}           & 0.460 & 0.483 & 0.418 & 0.463 & 0.684 \\
 \textbf{PointNet}      & \textbf{0.429} & \textbf{0.473} & \textbf{0.381} & \textbf{0.429} & \textbf{0.657} \\
 \textbf{GraphSAGE}     & 0.793 & 0.963 & 0.657 & 0.845 & 1.857 \\
 \textbf{GUNet}         & 0.745 & 0.859 & 0.679 & 0.742 & 0.996 \\

 \hline
\end{tabular}}\par
} 
\label{tab:5f}
\end{table}

\begin{table}
\caption{Table of RMSEs produced from models training and testing on 20 aerofoils}
{
\centering
\resizebox{\columnwidth}{!}{\begin{tabular}{ |c||c c|c c c|  }
 \hline
 \multicolumn{6}{|c|}{\textbf{20 Aerofoils}} \\
 \hline
Model & \multicolumn{2}{c|}{Train RMSE}  & \multicolumn{3}{c|}{Test RMSE} \\
& Surface & Fluid & Surface & Fluid & CL \\
 \hline
 \textbf{MLP}          & 0.587 & 0.297 &  \textbf{0.788} & \textbf{0.311} & \textbf{0.997}  \\
 \textbf{PointNet}     & 0.747 & 0.396 &  0.881 & 0.358 & 1.015  \\
 \textbf{GraphSAGE}    & \textbf{0.563} & \textbf{0.288} &  1.820 & 1.024 & 1.394  \\
 \textbf{GUNet}        & 0.578 & 0.306 &  1.808 & 0.653 & 1.335  \\
 \hline
\end{tabular}}\par
}
{
\centering
\resizebox{\columnwidth}{!}{\begin{tabular}{ |c||c c c c c|  }
 \hline
 \multicolumn{6}{|c|}{\textbf{Test RMSE Per Variable at Surface}} \\
 \hline
 Model & $\rho$&$\sigma_u$&$\sigma_v$&$e$&$\omega$ \\
 \hline
 \textbf{MLP}           & \textbf{0.599} & \textbf{0.222} & \textbf{0.124} & \textbf{0.658} & \textbf{1.498} \\
 \textbf{PointNet}      & 0.602 & 0.470 & 0.249 & 0.664 & 1.673 \\
 \textbf{GraphSAGE}     & 1.262 & 0.599 & 0.472 & 1.404 & 3.523 \\
 \textbf{GUNet}         & 1.097 & 0.625 & 0.454 & 1.249 & 3.604 \\

 \hline
\end{tabular}}\par
}
{
\centering
\resizebox{\columnwidth}{!}{\begin{tabular}{ |c||c c c c c|  }
 \hline
 \multicolumn{6}{|c|}{\textbf{Test RMSE Per Variable in Fluid}} \\
 \hline
 Model & $\rho$&$\sigma_u$&$\sigma_v$&$e$&$\omega$ \\
 \hline
 \textbf{MLP}           & 0.359 & \textbf{0.317} & \textbf{0.243} & \textbf{0.352} & \textbf{0.267} \\
 \textbf{PointNet}      & \textbf{0.354} & 0.365 & 0.271 & 0.353 & 0.429 \\
 \textbf{GraphSAGE}     & 0.750 & 0.741 & 0.720 & 0.769 & 1.737 \\
 \textbf{GUNet}         & 0.633 & 0.638 & 0.515 & 0.628 & 0.817 \\

 \hline
\end{tabular}}\par
} 

\label{tab:20f}
\end{table}
\begin{table}
\caption{Table of RMSEs produced from models training and testing on 55 aerofoils}
{
\centering
\resizebox{\columnwidth}{!}{\begin{tabular}{ |c||c c|c c c|  }
 \hline
 \multicolumn{6}{|c|}{\textbf{55 Aerofoils}} \\
 \hline
Model & \multicolumn{2}{c|}{Train RMSE}  & \multicolumn{3}{c|}{Test RMSE} \\
& Surface & Fluid & Surface & Fluid & CL \\
 \hline
 \textbf{MLP}          & 0.560 & 0.217 &  \textbf{0.729} & \textbf{0.261} & 0.795   \\
 \textbf{PointNet}     & 0.638 & 0.243 &  0.764 & 0.268 & \textbf{0.748}   \\
 \textbf{GraphSAGE}    & \textbf{0.472} & \textbf{0.191} &  1.543 & 0.760 & 1.299   \\
 \textbf{GUNet}        & 0.617 & 0.228 &  1.830 & 0.615 & 1.493   \\
 \hline
\end{tabular}}\par
}{
\centering
\resizebox{\columnwidth}{!}{\begin{tabular}{ |c||c c c c c|  }
 \hline
 \multicolumn{6}{|c|}{\textbf{Test RMSE Per Variable at Surface}} \\
 \hline
 Model & $\rho$&$\sigma_u$&$\sigma_v$&$e$&$\omega$ \\
 \hline
 \textbf{MLP}           & \textbf{0.551} & \textbf{0.115} & \textbf{0.072} & 0.610 & \textbf{1.402} \\
 \textbf{PointNet}      & 0.544 & 0.157 & 0.111 & \textbf{0.602} & 1.492 \\
 \textbf{GraphSAGE}     & 1.011 & 0.397 & 0.602 & 1.129 & 3.015 \\
 \textbf{GUNet}         & 1.161 & 0.380 & 0.328 & 1.311 & 3.663 \\

 \hline
\end{tabular}}\par
}
{
\centering
\resizebox{\columnwidth}{!}{\begin{tabular}{ |c||c c c c c|  }
 \hline
 \multicolumn{6}{|c|}{\textbf{Test RMSE Per Variable in Fluid}} \\
 \hline
 Model & $\rho$&$\sigma_u$&$\sigma_v$&$e$&$\omega$ \\
 \hline
 \textbf{MLP}           & 0.310 & \textbf{0.247} & \textbf{0.198} & 0.304 & \textbf{0.228} \\
 \textbf{PointNet}      & \textbf{0.308} & 0.265 & 0.209 & \textbf{0.301} & 0.243 \\
 \textbf{GraphSAGE}     & 0.617 & 0.712 & 0.736 & 0.605 & 1.046 \\
 \textbf{GUNet}         & 0.624 & 0.530 & 0.519 & 0.606 & 0.763 \\

 \hline
\end{tabular}}\par
} 

\label{tab:55f}
\end{table}
\begin{table}
\caption{Table of RMSEs produced from models training and testing on 150 aerofoils}
{
\centering
\resizebox{\columnwidth}{!}{\begin{tabular}{ |c||c c|c c c|  }
 \hline
 \multicolumn{6}{|c|}{\textbf{150 Aerofoils}} \\
 \hline
Model & \multicolumn{2}{c|}{Train RMSE}  & \multicolumn{3}{c|}{Test RMSE} \\
& Surface & Fluid& Surface & Fluid & CL \\
 \hline
 \textbf{MLP}          & \textbf{0.428} & \textbf{0.146} &  \textbf{0.335} & \textbf{0.138} & \textbf{0.861}  \\
 \textbf{PointNet}     & 0.586 & 0.186 &  0.458 & 0.171 & 0.954  \\
 \textbf{GraphSAGE}    & 0.472 & 0.153 &  1.946 & 0.913 & 1.168  \\
 \textbf{GUNet}        & 0.536 & 0.173 &  2.384 & 0.600 & 1.358  \\
 \hline
\end{tabular}}\par
}
{
\centering
\resizebox{\columnwidth}{!}{\begin{tabular}{ |c||c c c c c|  }
 \hline
 \multicolumn{6}{|c|}{\textbf{Test RMSE Per Variable at Surface}} \\
 \hline
 Model & $\rho$&$\sigma_u$&$\sigma_v$&$e$&$\omega$ \\
 \hline
 \textbf{MLP}           & \textbf{0.212} & \textbf{0.107} & \textbf{0.061} & \textbf{0.233} & \textbf{0.668} \\
 \textbf{PointNet}      & 0.284 & 0.158 & 0.102 & 0.315 & 0.913 \\
 \textbf{GraphSAGE}     & 1.318 & 0.292 & 0.313 & 1.467 & 3.856 \\
 \textbf{GUNet}         & 1.006 & 0.377 & 0.760 & 1.222 & 5.019 \\

 \hline
\end{tabular}}\par
}
{
\centering
\resizebox{\columnwidth}{!}{\begin{tabular}{ |c||c c c c c|  }
 \hline
 \multicolumn{6}{|c|}{\textbf{Test RMSE Per Variable in Fluid}} \\
 \hline
 Model & $\rho$&$\sigma_u$&$\sigma_v$&$e$&$\omega$ \\
 \hline
 \textbf{MLP}           & \textbf{0.136} & \textbf{0.159} & \textbf{0.121} & \textbf{0.137} & \textbf{0.131} \\
 \textbf{PointNet}      & 0.172 & 0.183 & 0.144 & 0.172 & 0.181 \\
 \textbf{GraphSAGE}     & 0.733 & 0.594 & 0.600 & 0.725 & 1.547 \\
 \textbf{GUNet}         & 0.444 & 0.436 & 0.401 & 0.447 & 1.025 \\

 \hline
\end{tabular}}\par
}
\label{tab:150f}
\end{table}
\begin{table}
\caption{The inference time (time taken per epoch) for each model}
{
\centering
\label{tab:time}
\resizebox{0.75\columnwidth}{!}{\begin{tabular}{ |c c c c|  }
 \hline
 \multicolumn{4}{|c|}{\textbf{Inference Time ($m s$)}} \\
 \hline
 \hline
 \textbf{MLP} &  \textbf{PointNet}  & \textbf{GraphSAGE}  &   \textbf{GUNet}  \\

 1.36 & 3.64 & 3.03 & 45.28 \\

 \hline
\end{tabular}}\par
}
\end{table}
\begin{figure*}
    
\begin{multicols}{2}
    \includegraphics[width=1\linewidth]{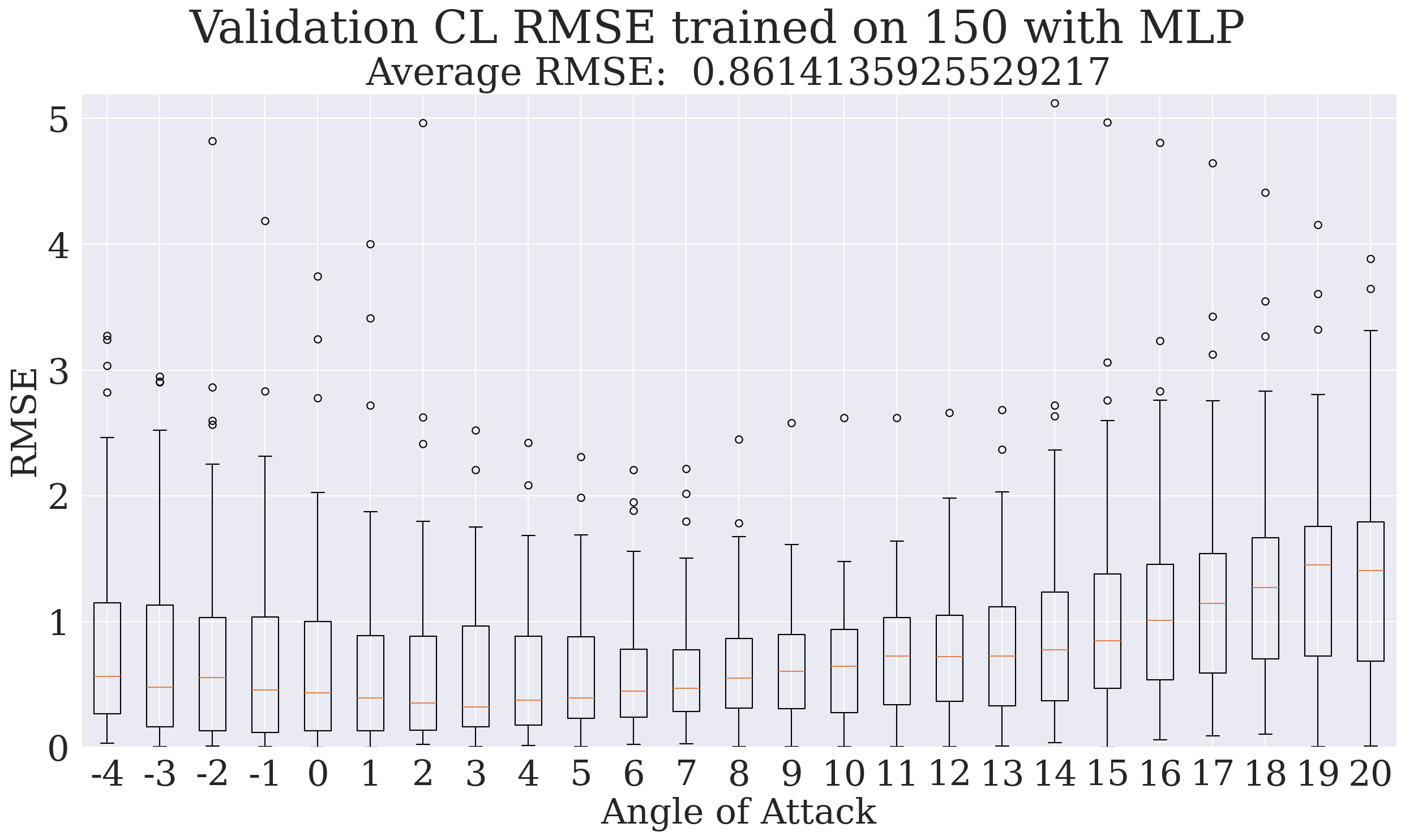}\par
    \includegraphics[width=1\linewidth]{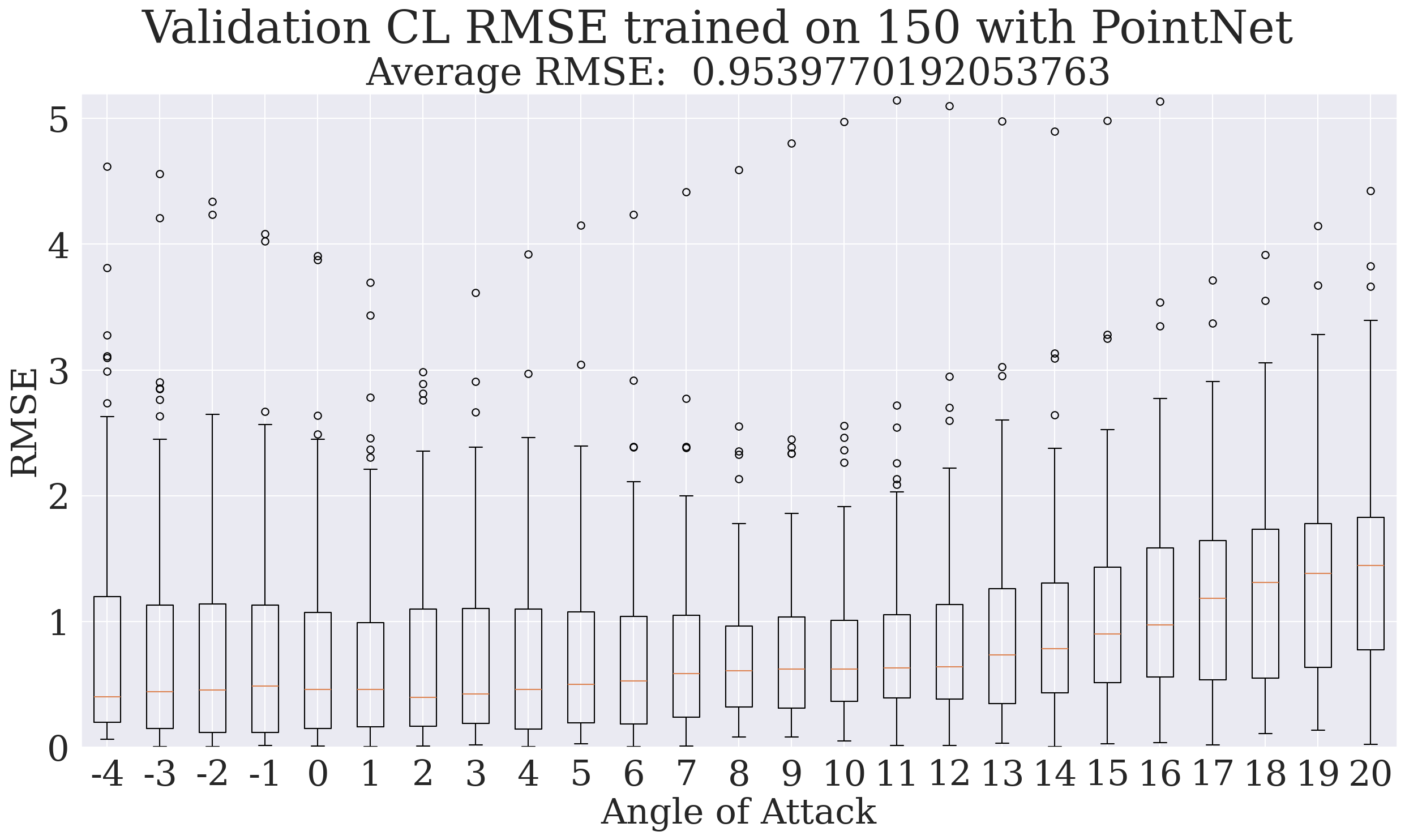}\par
    \end{multicols}
\begin{multicols}{2}
    \includegraphics[width=1\linewidth]{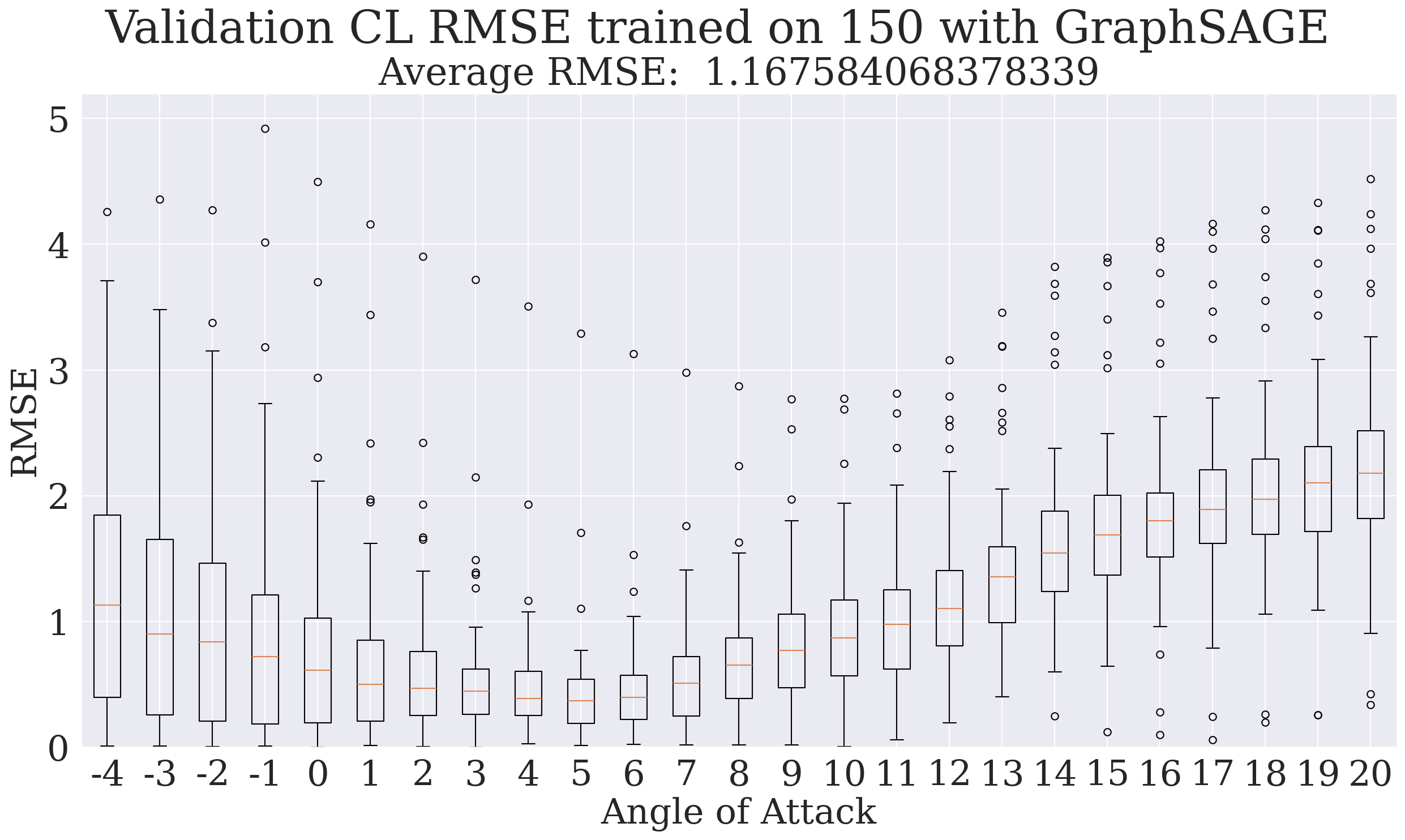}\par
    \includegraphics[width=1\linewidth]{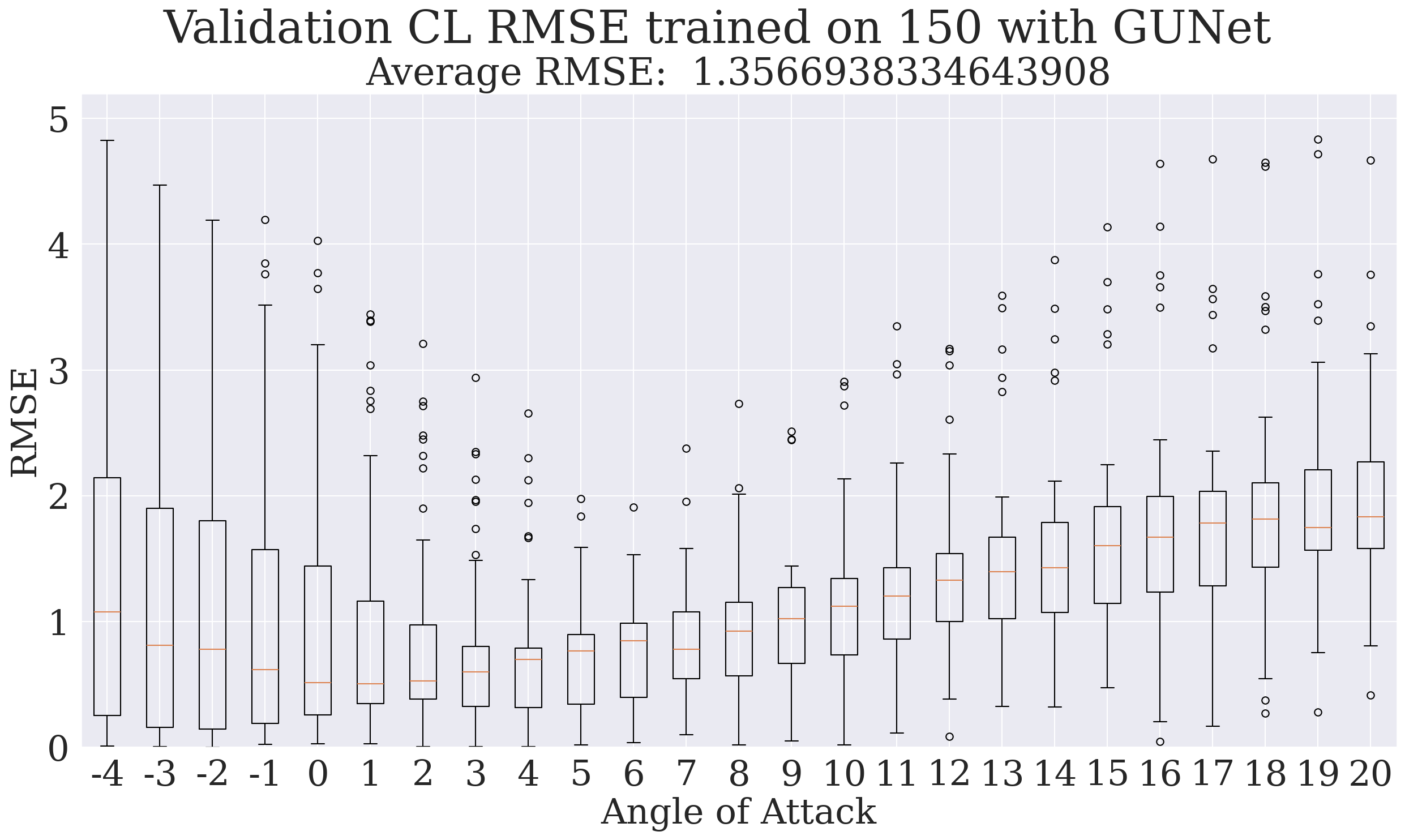}\par
    \end{multicols}
    \caption{Box Plots showing the RMSE between the calculated and real $C_L$ for all 60 validation aerofoils at each angle of attack trained on 150 aerofoils}
    \label{fig:box}
\end{figure*}

\begin{figure*}
    
\begin{multicols}{2}
    \includegraphics[width=1\linewidth]{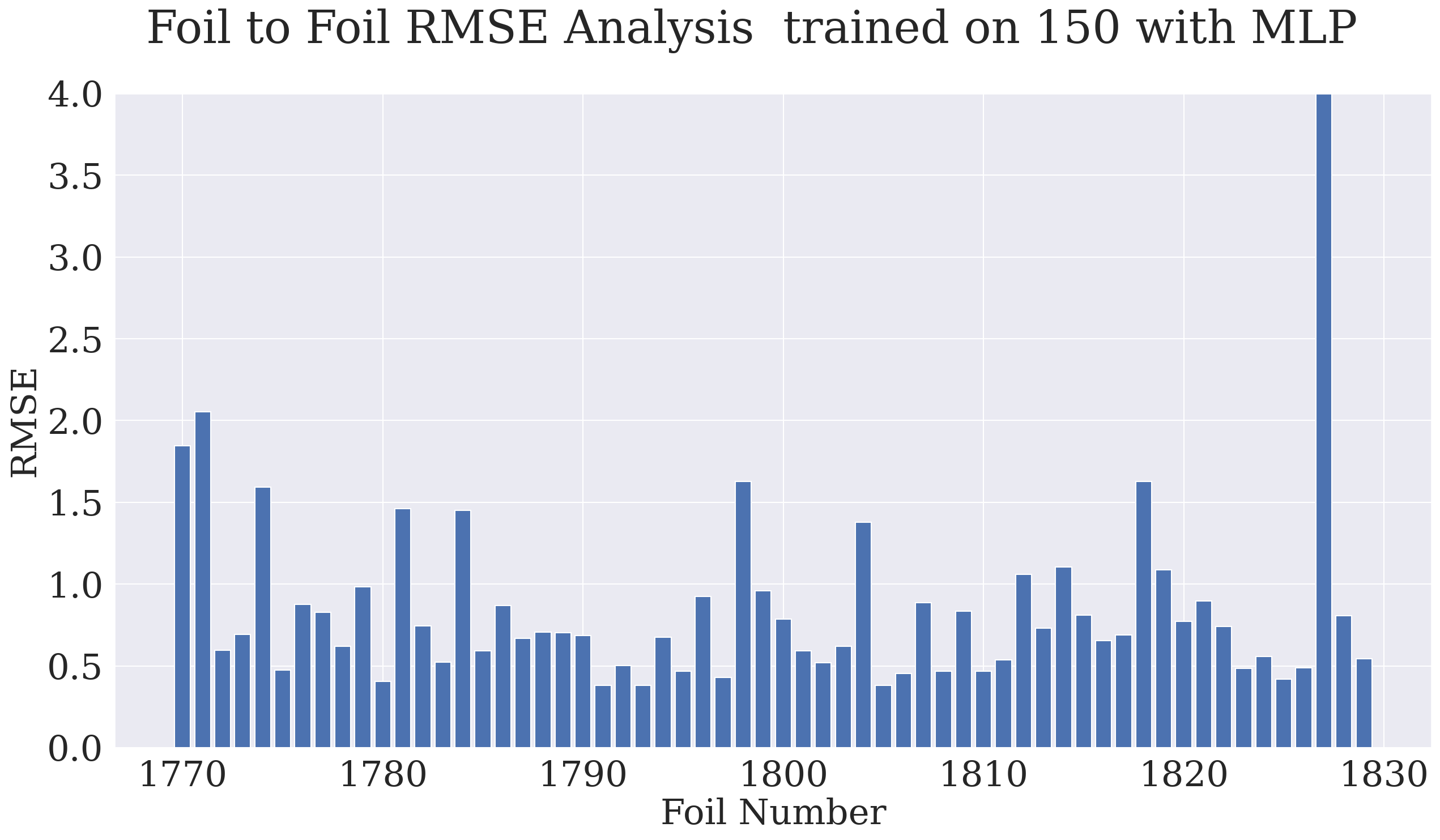}\par
    \includegraphics[width=1\linewidth]{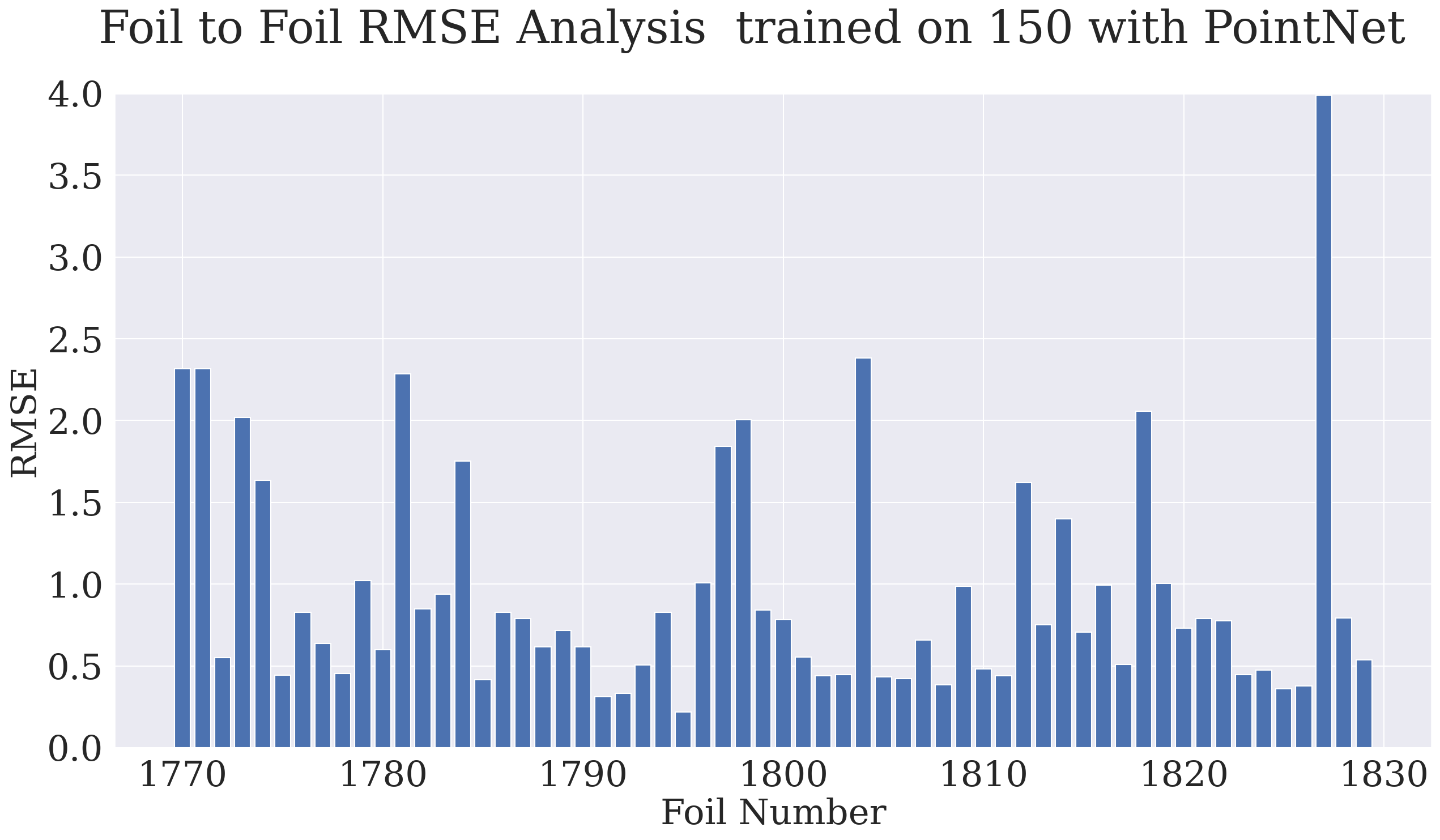}\par
    \end{multicols}
\begin{multicols}{2}
    \includegraphics[width=1\linewidth]{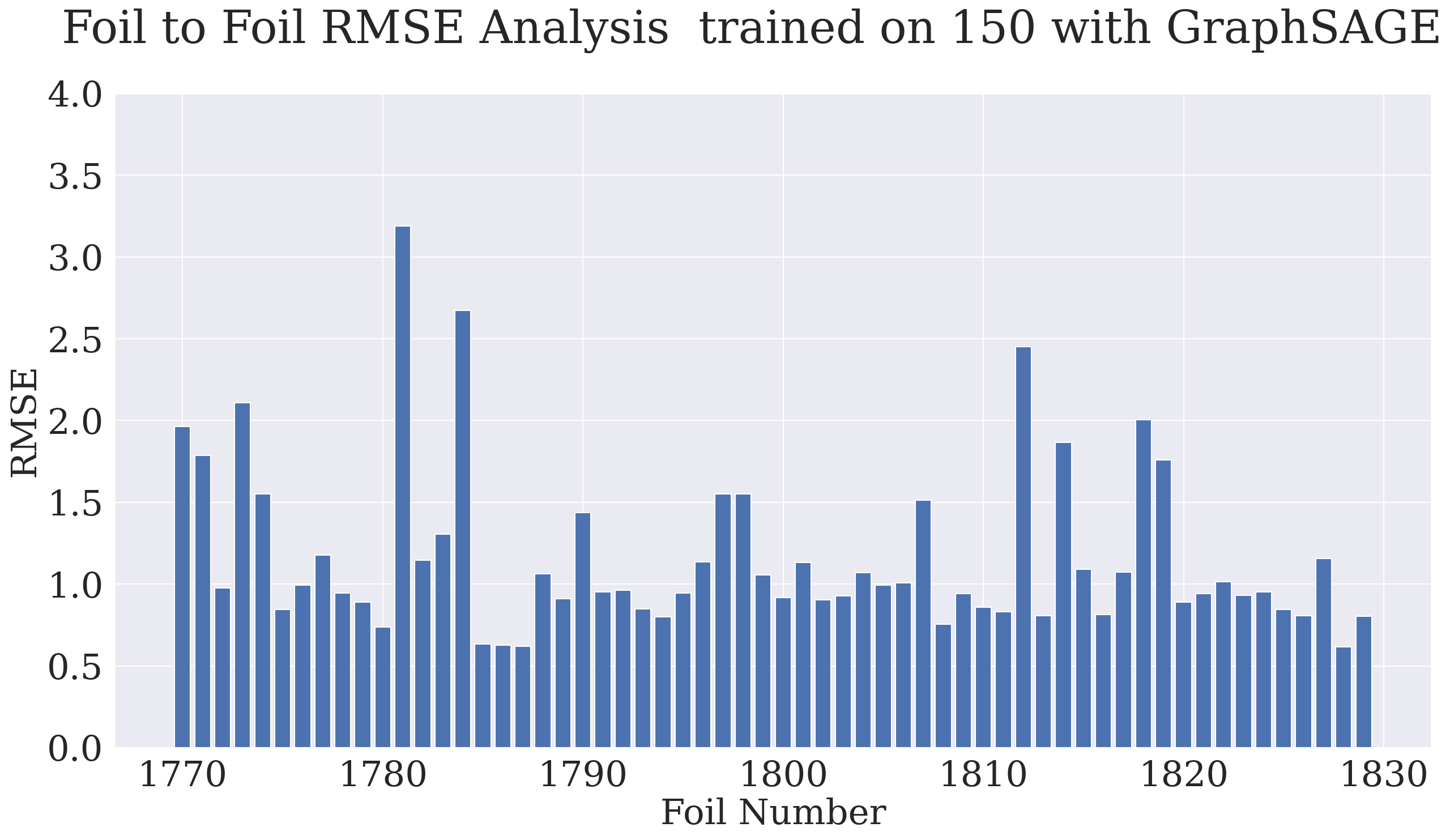}\par
    \includegraphics[width=1\linewidth]{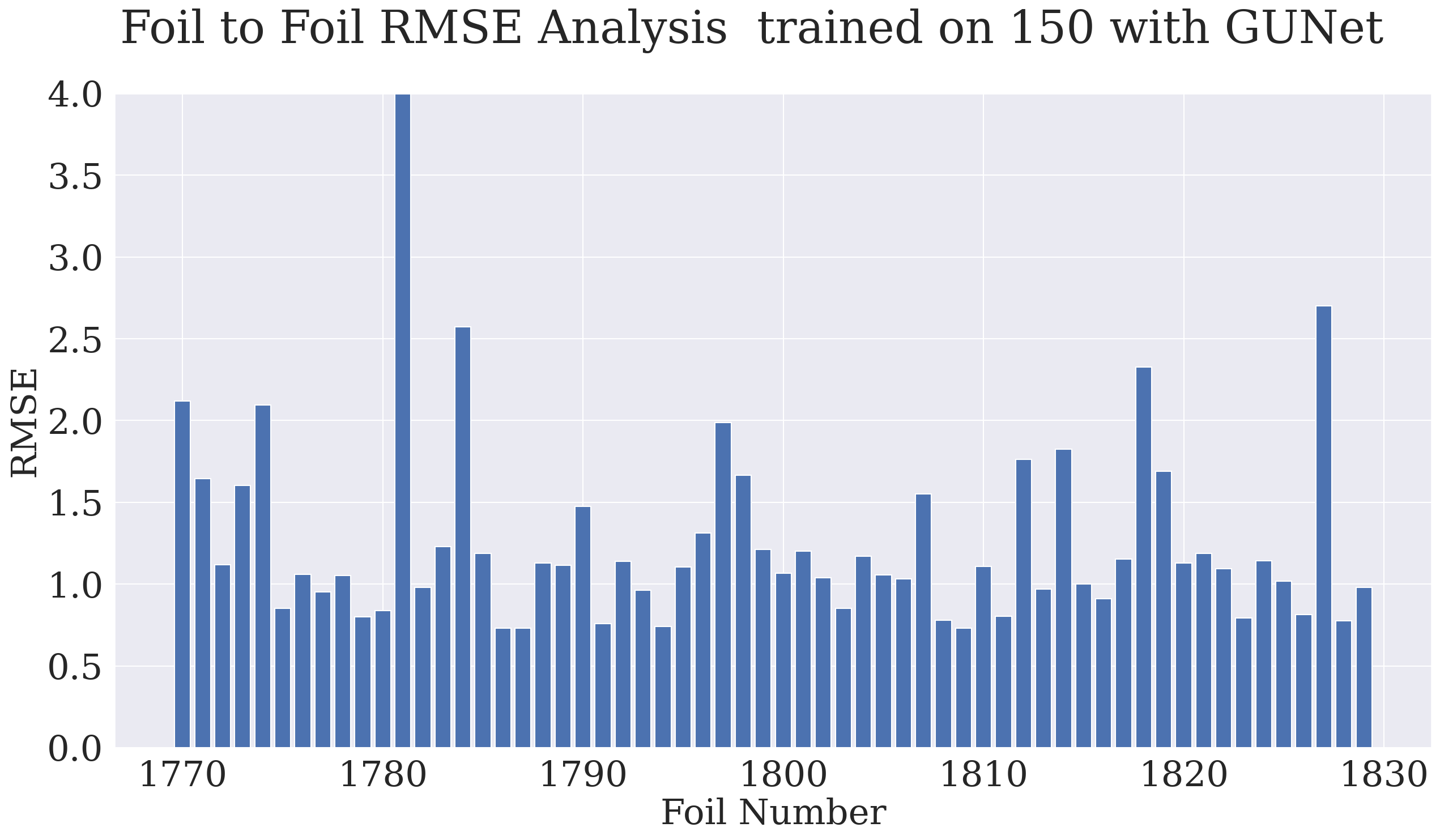}\par
    \end{multicols}
    \caption{Bar Chart showing the RMSE between the calculated and real $C_L$, averaged over every angle of attack for all 60 validation foils}
    \label{fig:foilbyfoil}
\end{figure*}

\section{Discussion}

There are a few key results to note from the outputs of the models, as shown in Tables \ref{tab:5f}-\ref{tab:150f}.
Firstly, despite their relative complexity and consistently high performance across the training regime, both the GraphSAGE and GUNet models achieve relatively high errors in the test set, regularly over double that of PointNet and MLP.
The likely reason for such behaviour is that both GraphSAGE and GUNet are over-fitting to the data in training; this means the models are memorising the data they are being trained on, instead of learning the patterns or trends within the data.

Table \ref{tab:5f} shows notably less correlation across the results than the following tables, with MLP and PointNet performing best over an equal number of metrics at the surface. This discrepancy when compared to the rest of the Tables, as well as the large RMSE across testing, indicates that 5 Aerofoils is insufficient to reliably predict the behaviour of the fluid within an acceptable margin of error.

Across the overall test metrics, MLP predicts the values at the surface and the fluid the most accurately seven out of eight times, coming in second to PointNet only in Table \ref{tab:5f}. Interestingly, in Table \ref{tab:55f} despite being slightly less accurate at predicting $\omega$ at the surface, PointNet predicts a more accurate $C_L$. 
Tables \ref{tab:55f} and \ref{tab:150f}, however, show the expected results, with the model most accurately able to predict the vorticity at the surface, resulting in the most accurate $C_L$.

\begin{figure}[h]
    \centering
    \includegraphics[width=1\linewidth]{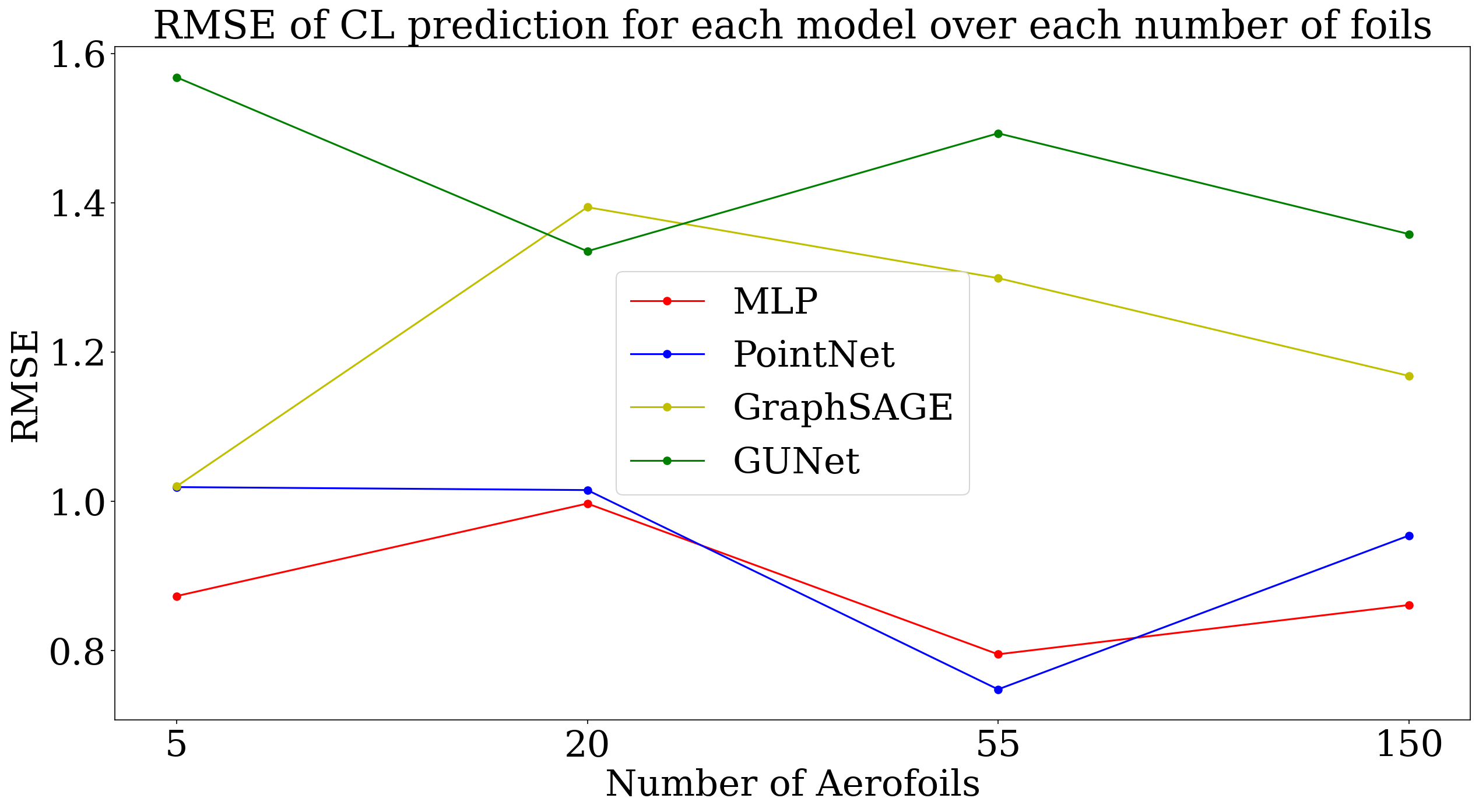}
    \caption{Comparison of the RMSE of $C_L$ predicted by each model against how many aerofoils the models were trained upon}
    \label{fig:cl_per_foil}
\end{figure}

\begin{figure}[h]
    \centering
    \includegraphics[width=1\linewidth]{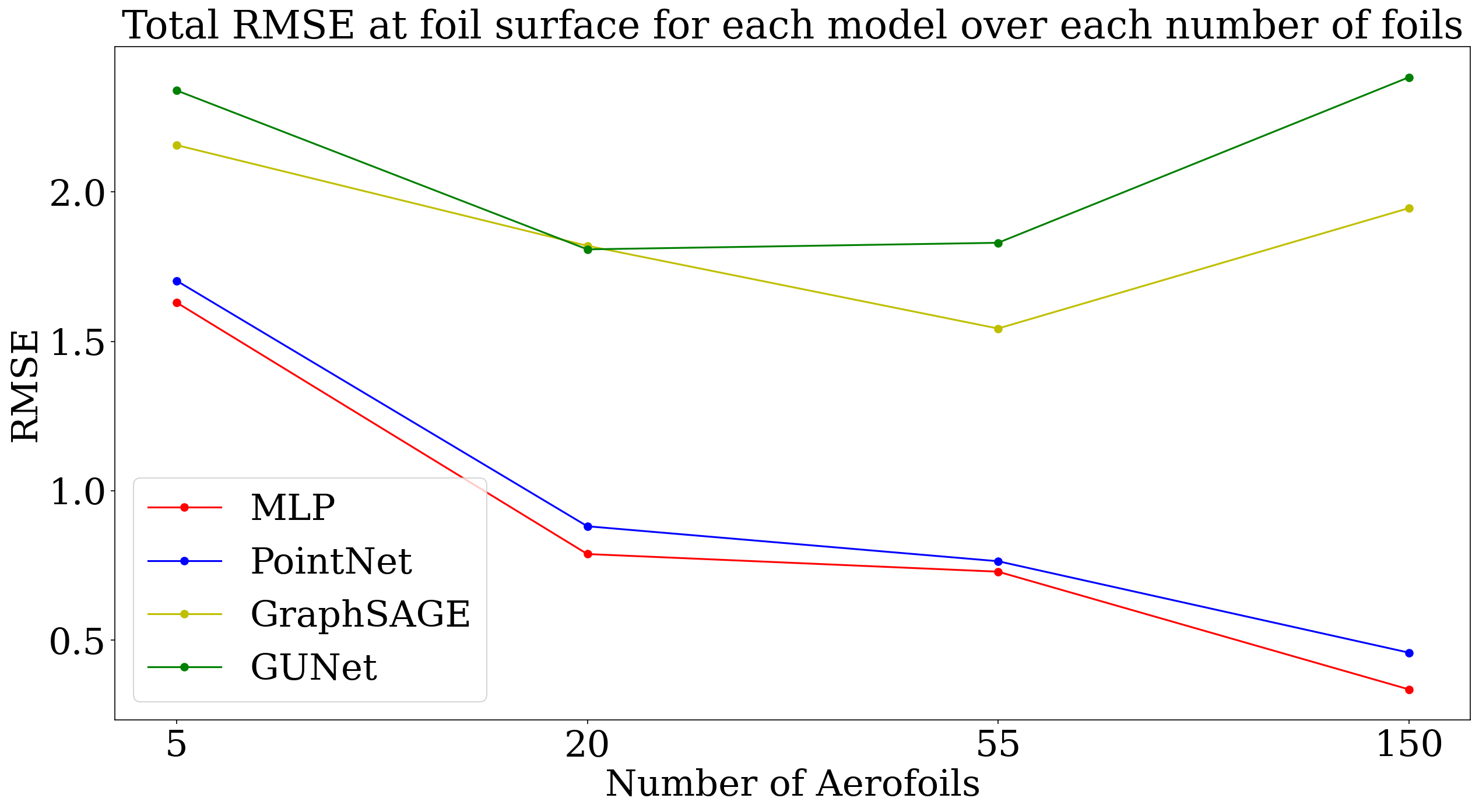}
    \caption{Comparison of the RMSE at the surface of the aerofoils predicted by each model against how many aerofoils the models were trained upon}
    \label{fig:rmse_surf}
\end{figure}

As evidenced in Fig. \ref{fig:cl_per_foil}, and Tables \ref{tab:5f}-\ref{tab:150f}, none of the aerofoil numbers show a consistent decrease in error for the prediction of $C_L$.
Interestingly, the point of lowest RMSE for $C_L$, across all experiments, is for PointNet using 55 aerofoils, despite the RMSE for $\omega$ being higher than that of MLP in Table \ref{tab:20f}. Notably, in Table \ref{tab:55f}, MLP has the lowest total RMSE at both the surface and in the fluid, aside for $C_L$. Furthermore, despite the test RMSEs at the surface of the foils decreasing between Tables \ref{tab:55f} and \ref{tab:150f}, the error for the calculation of $C_L$ actually increases.
Whilst the precise reasoning for this requires further research, the most likely reason for this is the simplified nature of the panel method as it relies solely on $\omega$ at the surface and the flow conditions at infinity.
Furthermore, the error for $\omega$ is the average over the whole of the surface of the foil, and larger errors at specific positions could have a larger effect of the overall calculation of $C_L$.

GraphSAGE and GUNet both show notably higher errors in both the prediction of variables at the foil surface as well as in the calculations of $C_L$. As can be seen in Fig. \ref{fig:rmse_surf}, in comparison with MLP and PointNet which show a decrease in surface RMSE with every increase in foil number, the graph methods show increases in error at 150 foils. 
These results combined with the longer inference required due to the more complicated nature of the models (as shown in Table \ref{tab:time}) indicate a lack of suitability of GUNet and GraphSAGE for this problem statement. Whilst they train well, all evidence points to the models over-fitting to the training data.

\begin{figure}[h]
    \centering
    \includegraphics[width=1\linewidth]{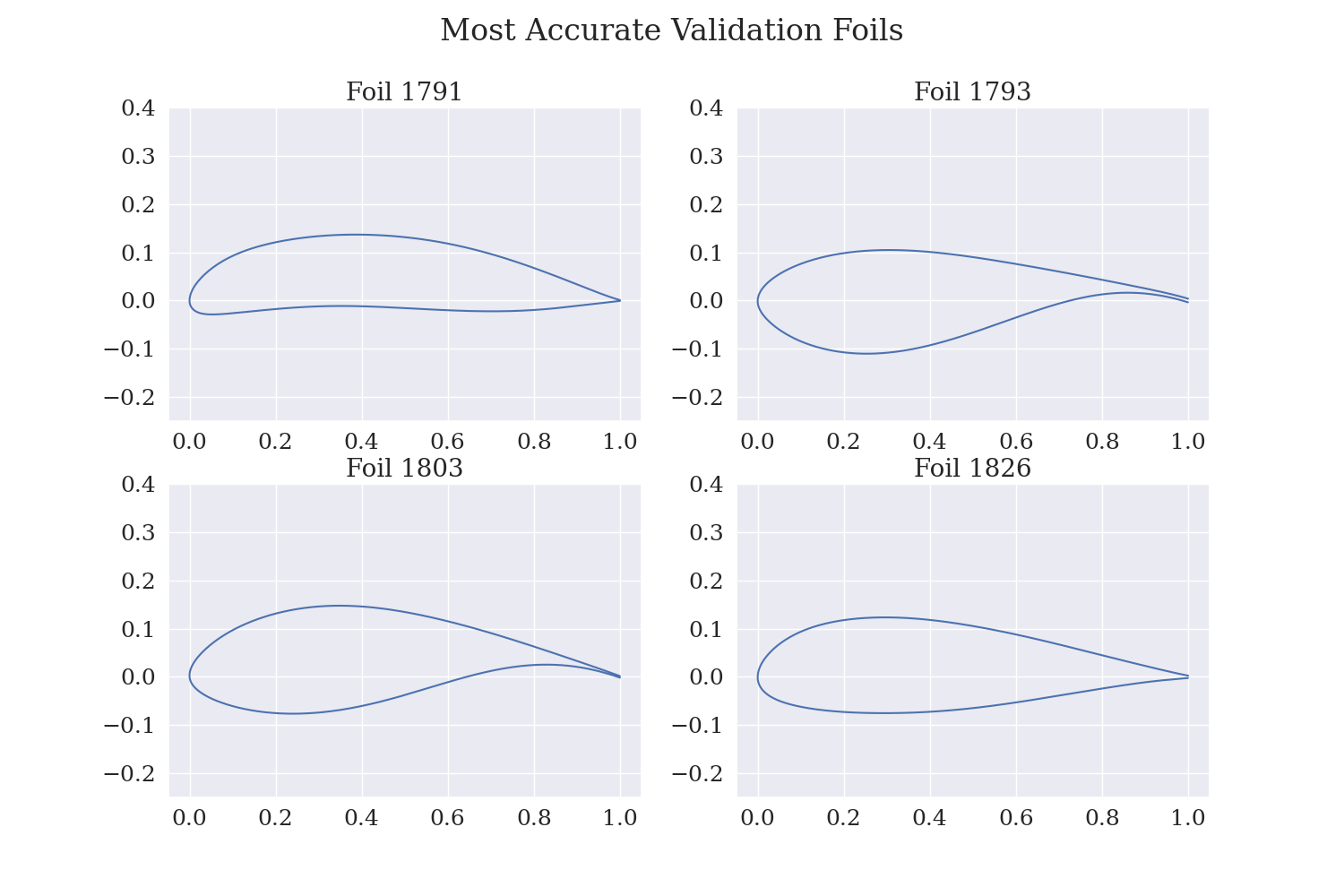}
    \caption{The four aerofoils with the lowest RMSEs across the models when trained on 150 aerofoils}
    \label{fig:Best}
\end{figure}

\begin{figure}[h]
    \centering
    \includegraphics[width=1\linewidth]{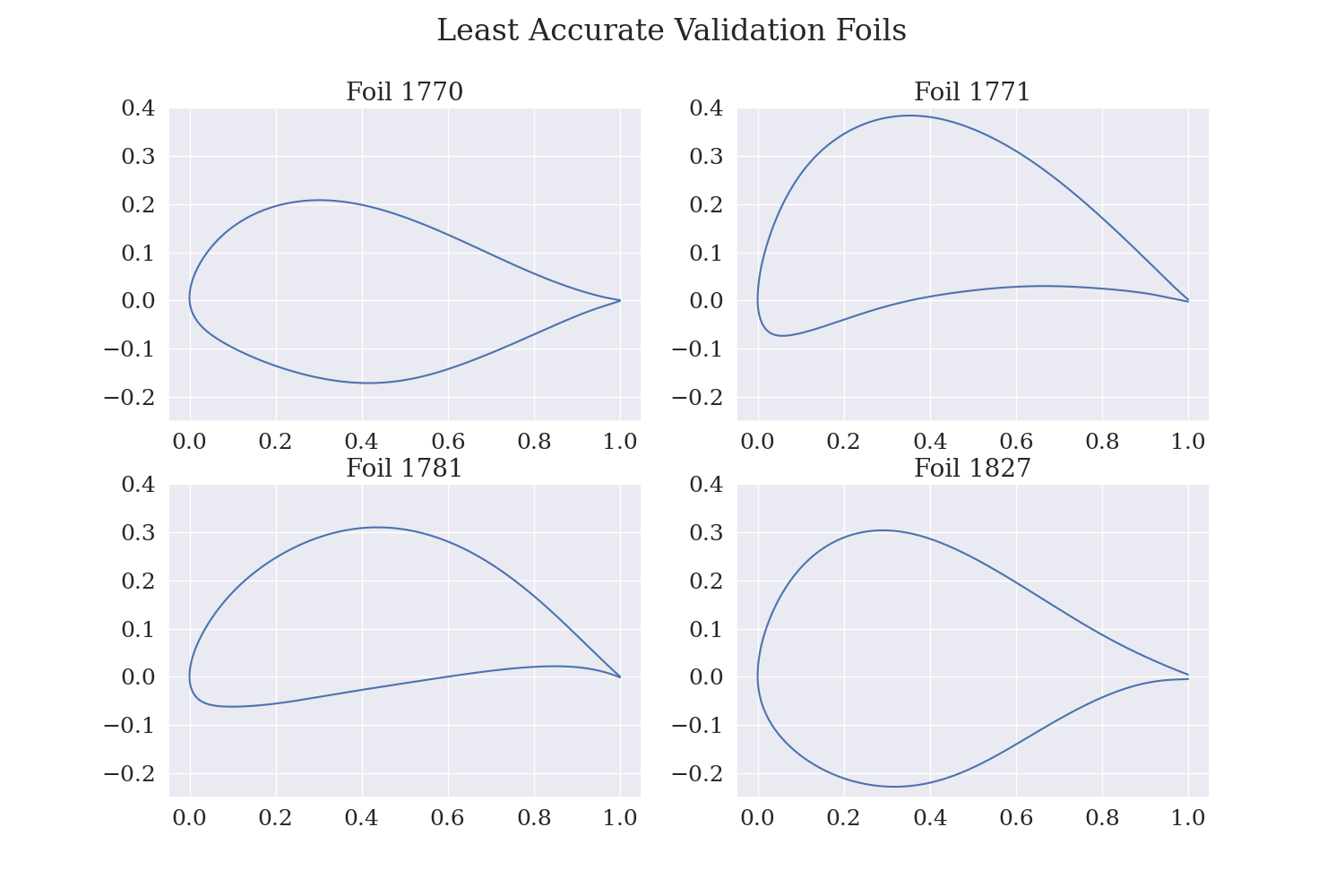}
    \caption{The four aerofoils with the highest RMSEs across the models when trained on 150 aerofoils}
    \label{fig:Worst}
\end{figure}

The foil-to-foil comparison, shown in Fig. \ref{fig:foilbyfoil}, provided further insight into the effect that the physical properties of each aerofoil had on the predictions of the models. The geometries of the most and least accurate four foils can be seen in Figs. \ref{fig:Best} and \ref{fig:Worst}, respectively. When the test aerofoils were compared to each other, four foils clearly stood out as having the highest errors across the models, whilst there were a large number of foils that resulted in relatively low RMSE. 


\begin{figure}[h]
    \includegraphics[width=.49\linewidth]{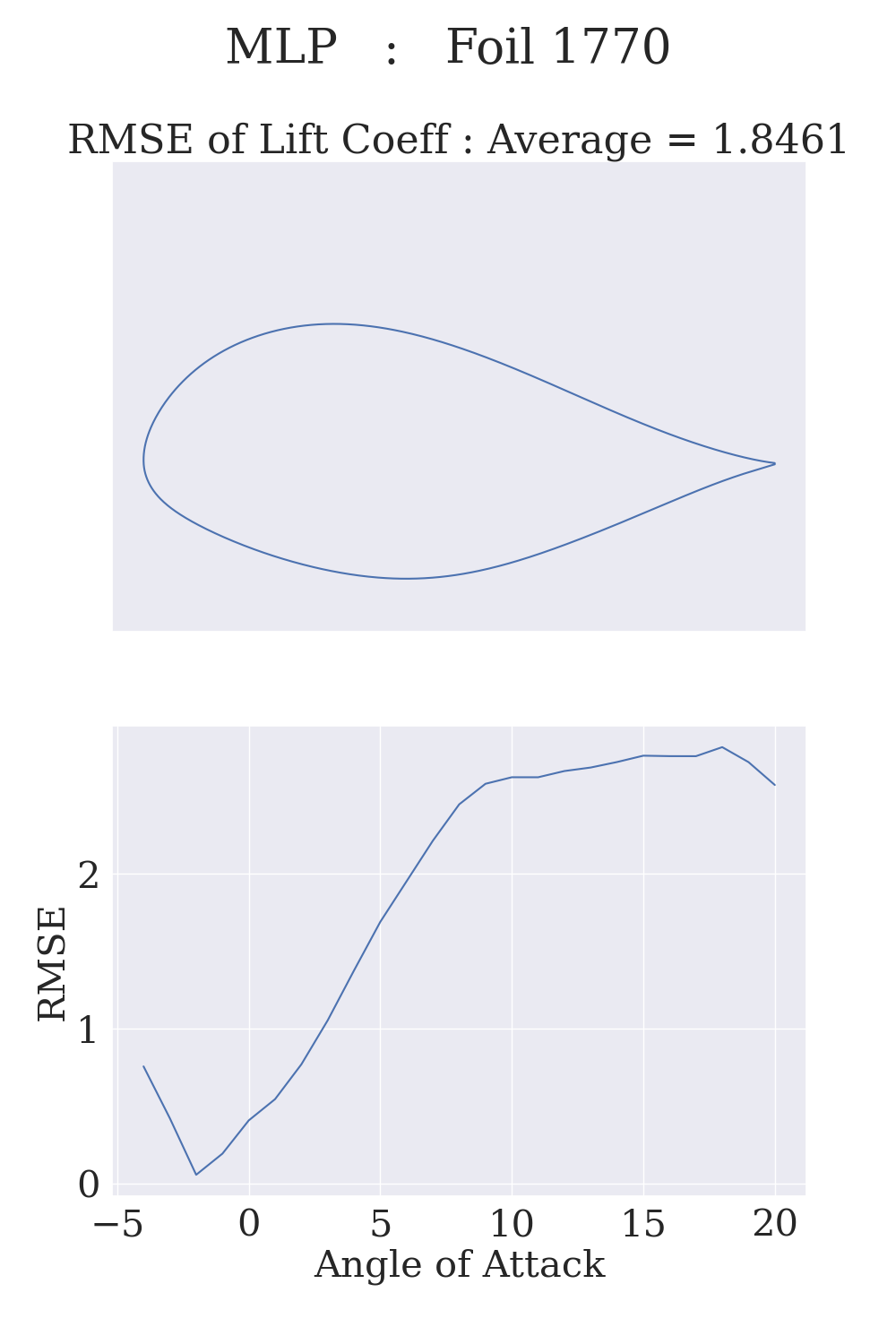}
    \includegraphics[width=.49\linewidth]{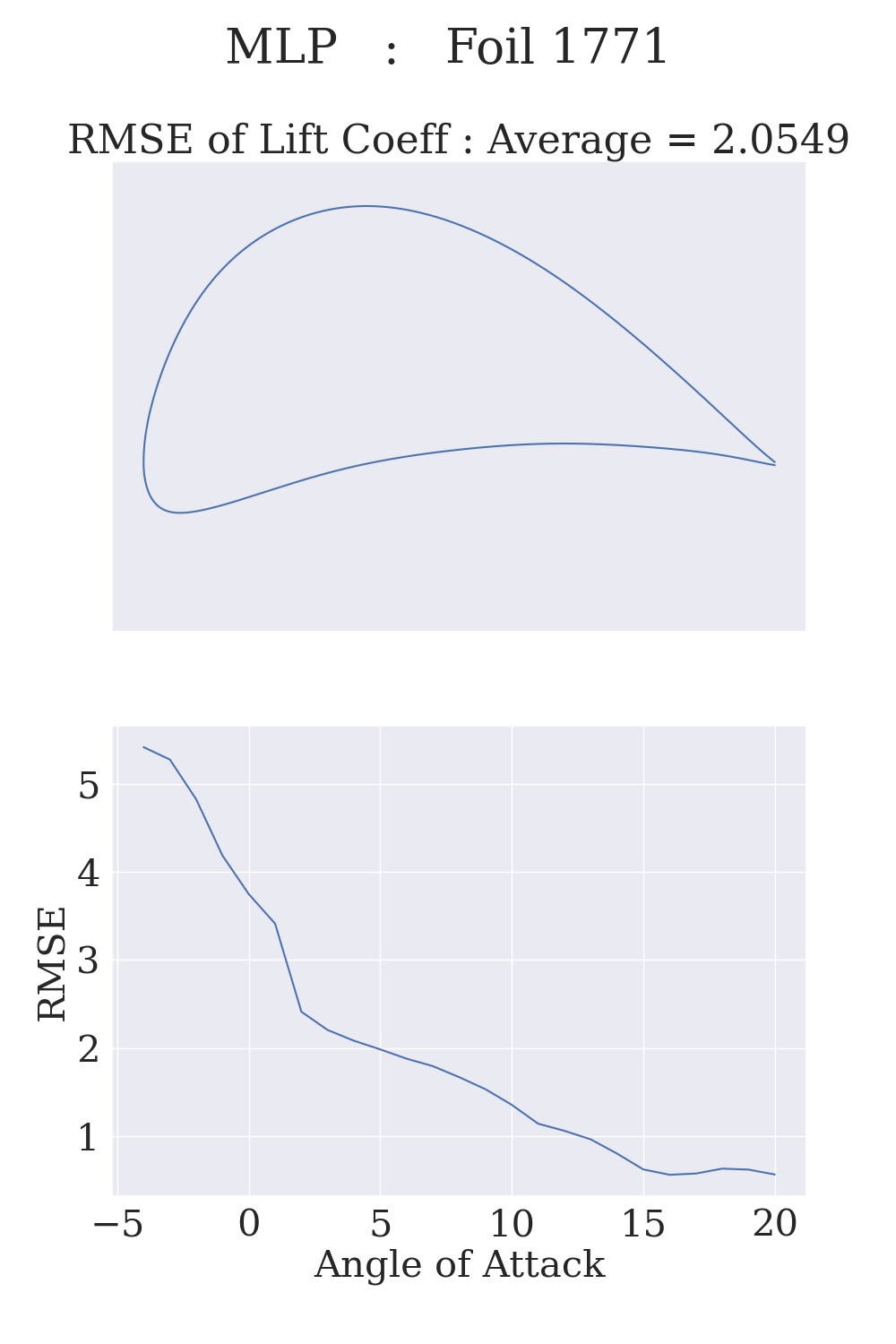}
    \includegraphics[width=.49\linewidth]{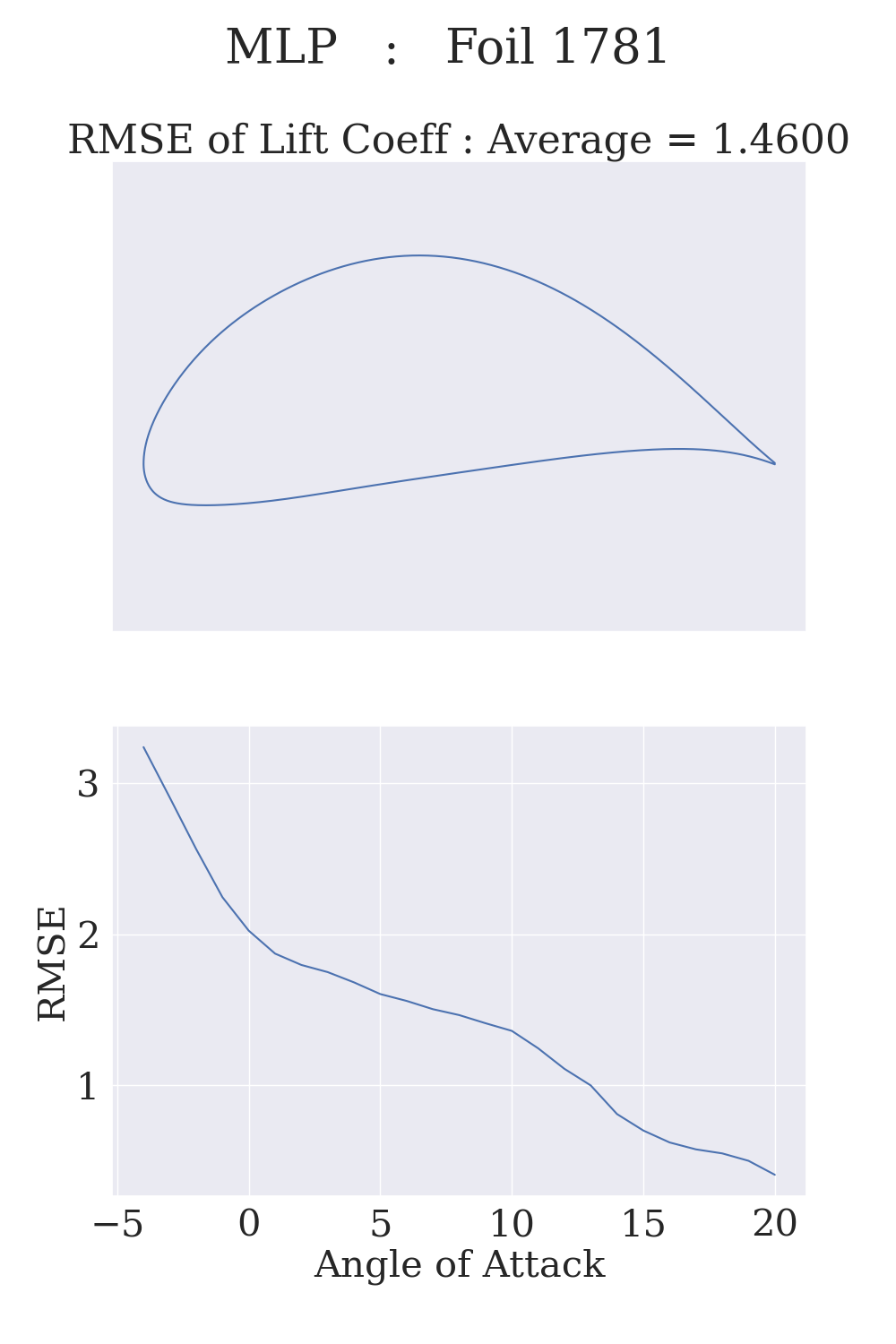}
    \includegraphics[width=.49\linewidth]{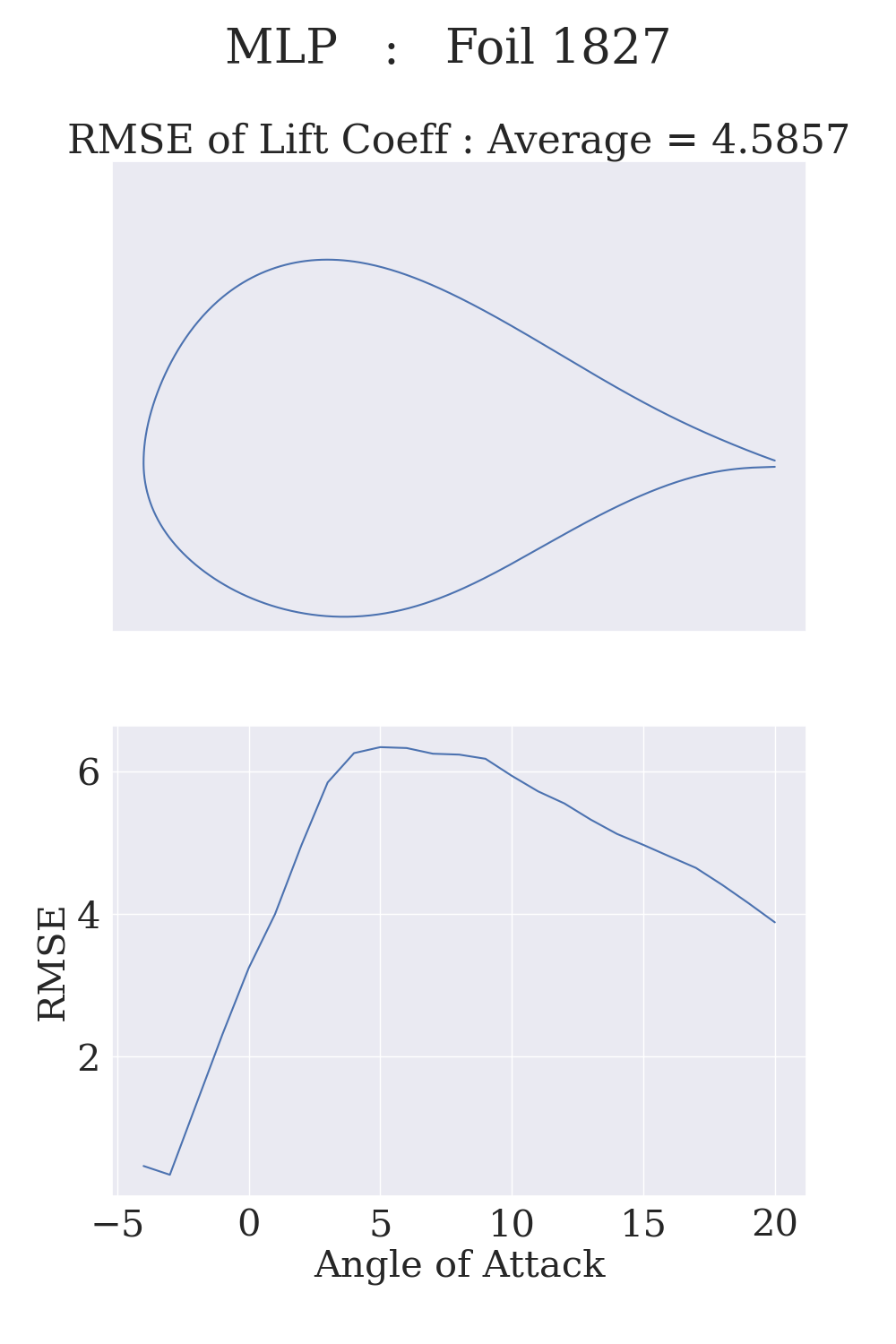}
    \caption{The four least accurate aerofoils shape and RMSE for $C_L$ across all four models acquired using MLP trained on 150 foils}
    \label{fig:foilVrmsepoor}
\end{figure}

\begin{figure}[h]
    \includegraphics[width=.49\linewidth]{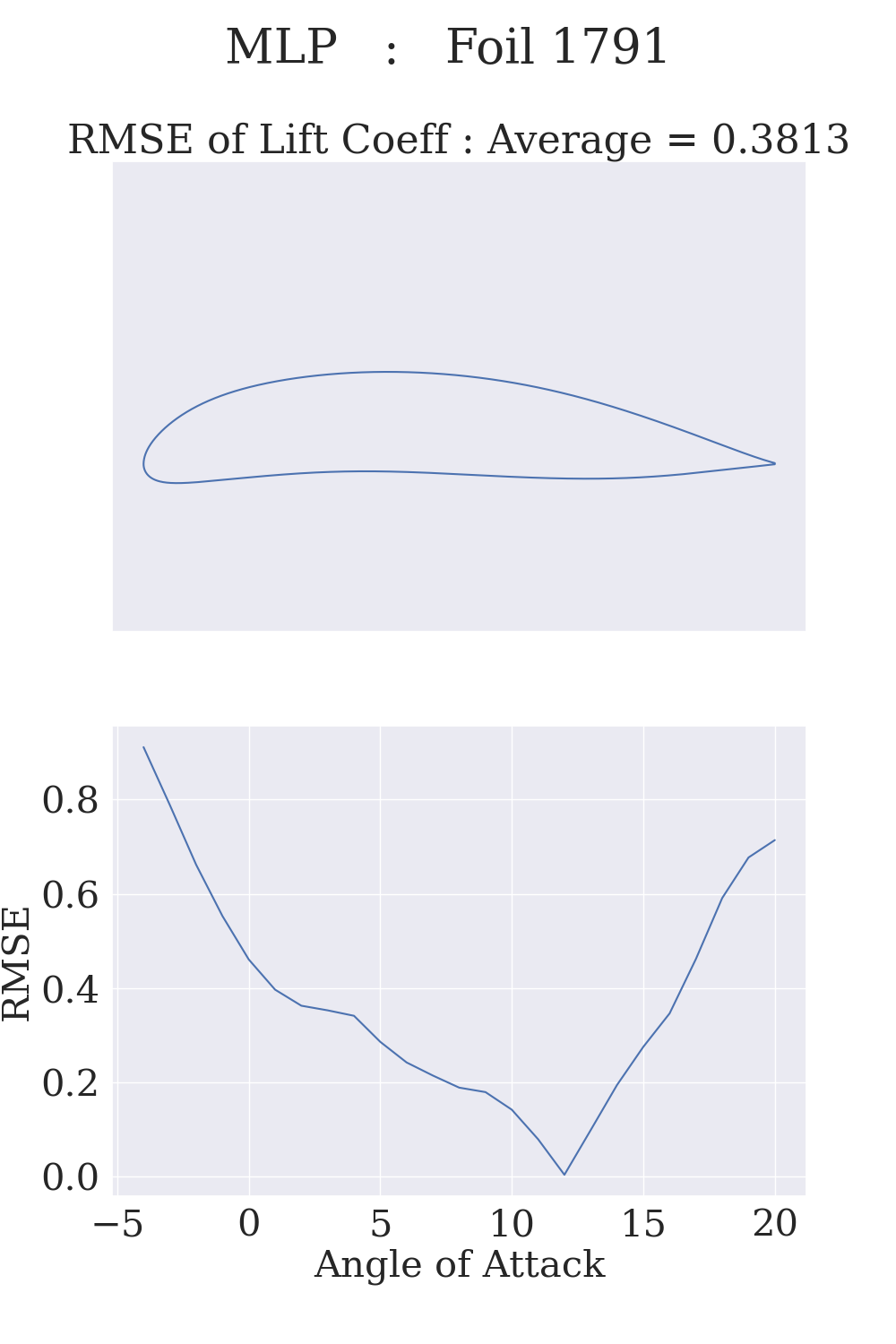}
    \includegraphics[width=.49\linewidth]{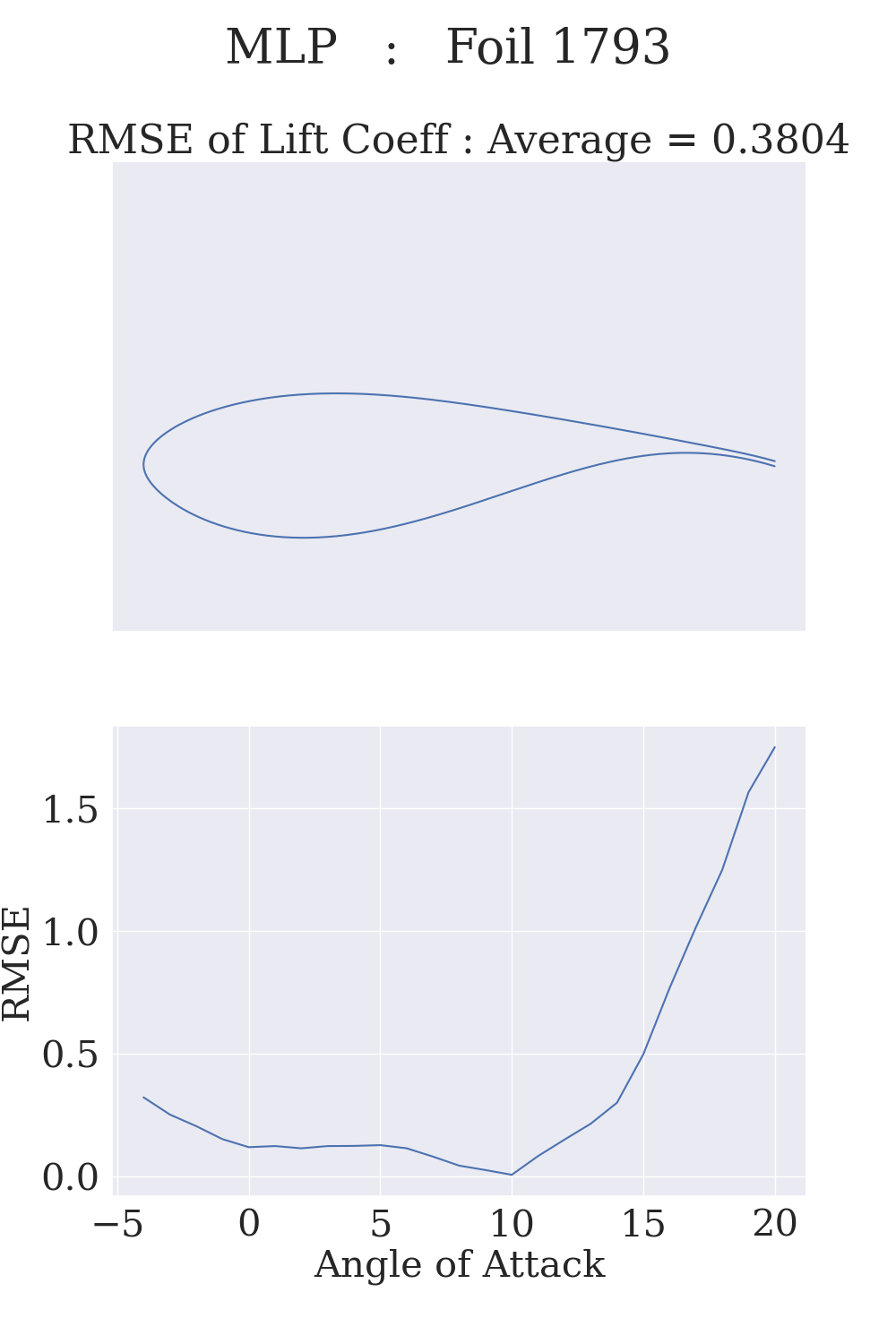}
    \includegraphics[width=.49\linewidth]{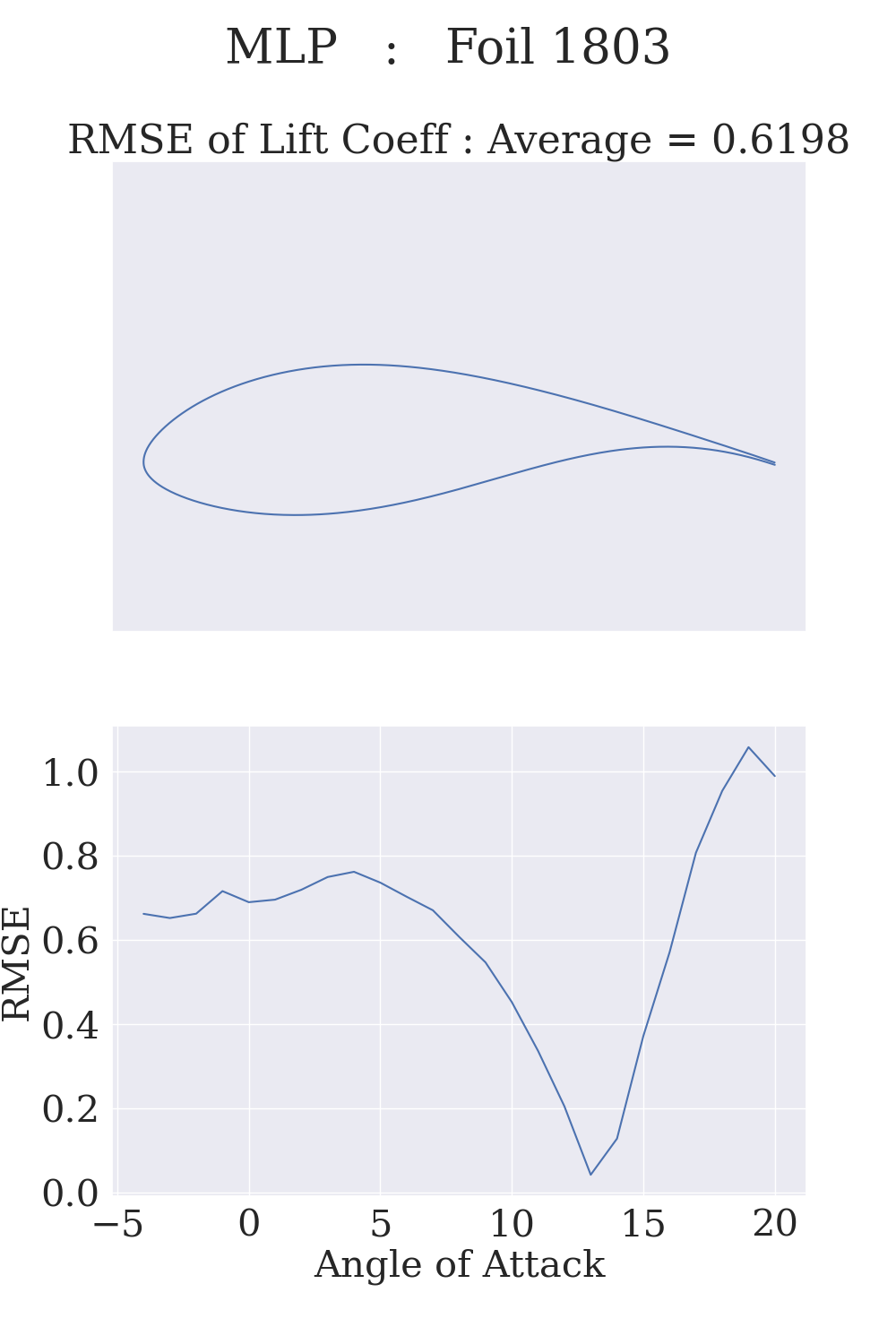}
    \includegraphics[width=.49\linewidth]{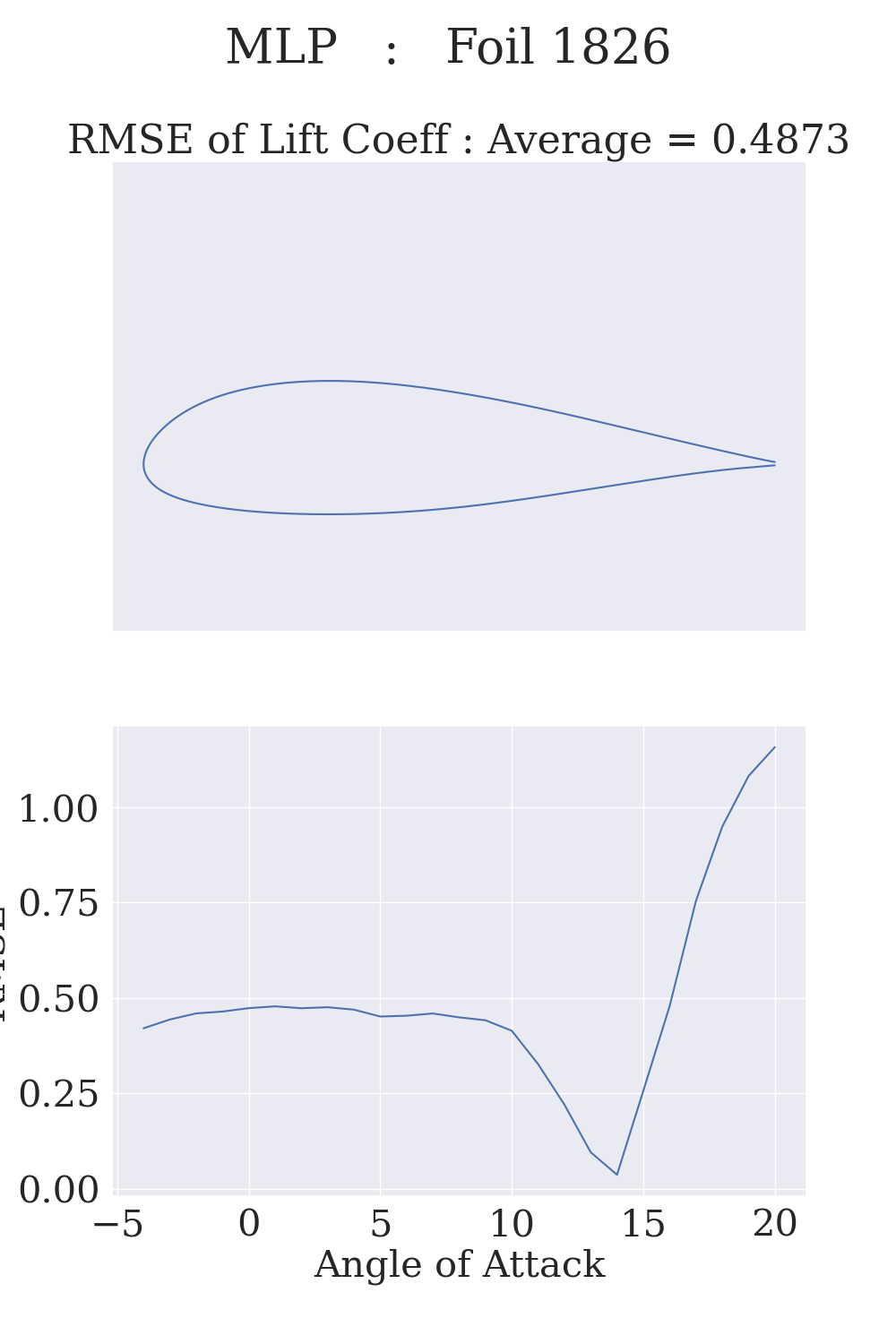}
    \caption{The four most accurate aerofoils shape and RMSE for $C_L$ across all four models acquired using MLP trained on 150 foils}
    \label{fig:foilVrmsegood}
\end{figure}

The differences between the two set of aerofoils are clear to see, the foils that performed better are thinner and have less of a camber. When comparing the RMSEs at each AoA for the less accurate foils, as shown in Fig. \ref{fig:foilVrmsepoor}, there is a visible correlation between the shape of the foils and whether the model was more accurate at low or high AoAs. The foils that were more rounded, with a larger lower camber were more accurate at lower angles of attack, whilst the foils with a larger mean camber were more accurate for higher angles of attack. Whereas the blades with a lower average RMSE for $C_L$ in Fig. \ref{fig:foilVrmsegood} are all less thick. Interestingly, the better performing blades all have points of lowest RMSE between 10$^\circ$ and 15$^\circ$ before having their highest RMSE at 20$^\circ$. 

Zhang et al. \cite{ZHANG201560} found that a `drastically varying surface curvature in the leading edge' caused early flow detachments along the surface of the aerofoil.
This separation causes the flow to reverse direction~\cite{ZHANG201560} which is more difficult for the models to predict than a simple and cohesive flow behaviour.

The relative thickness of aerofoil also plays a role in the accuracy of the model. Zhang et al.~\cite{MA20151003} found that, at low Re, a larger relative thickness causes a larger laminar separation bubble and affects the location of flow separation. Both of these factors effect the overall performance of the foil, as well as the models ability to predict the required physical parameters. Further to this, the aerofoils themselves have been defined in this dataset as a set of discrete points, which will introduce errors into the predictions. Whilst detrimental to the output of the results when compared to real world testing, this limitation is one that is shared with CFD~\cite{Spalart_Venkatakrishnan_2016} and as such should not affect the results listed in Section \ref{results}.

One minor issue found when analysing the aerofoil from the dataset is that none of the data points include the name of the aerofoil, making it difficult to verify the split of foils the model is being trained on. The model could, for example, be getting trained on a disproportionate number of NACA foils which would cause a lesser quality of generalisation to unseen, non-NACA aerofoils.

Unsurprisingly, the predictions generated by the models are somewhat compromised by the implementation of more extreme aerofoil designs. If NNs are to be a viable tool in the design process of aerofoils, they need to be able to generalise to any foil they are given robustly. Common practice is simply to use more data, but, as highlighted earlier, this is not an effective strategy given the lack of data available and the intense computational load that NNs require. 

As stated in Section \ref{LitRev}, Bonnet et al. \cite{AirfRANS} provided a benchmark to data similar to that used for this paper, with the main differences being the variables calculated. Bonnet et al. also used the models to predict $C_L$ and $C_D$ directly, whereas this paper calculated them using the predicted outputs. Whilst the results are comparable for the fluid predictions, the largest difference in accuracy lies in the predictions of the coefficients. One of the biggest drawbacks to using the panel method for calculating the coefficients is that it cannot be used to calculate the drag force acting on the aerofoil. Therefore, in order to use \texttt{windAI\_bench} to find $C_L$ and $C_D$, the coefficients must be predicted using the models.

The primary issue with the comparison of the two benchmarks is the lack of testing regime present in the \texttt{AirfRANS} paper. As this paper is more interested in the ability of these models to train on the data and then use that learnt knowledge to predict over unseen blades. To compare the training results of both papers would not provide any useful insights into the performance of the models and, as such, said comparison has not been performed.

\section{Conclusion}

This paper presents a benchmark for the \texttt{windAI\_bench} dataset created by NREL~\cite{windAi_bench}. It compares the predictions made by four models using a range of sample sizes to determine the models' efficacy on smaller data sets. The results from the models were then used to analytically calculate the lift coefficients of the aerofoils at a number of AoA.

This paper found that whilst GraphSAGE and GUNet performed well during the training phase, they performed poorly during testing, meaning their ability to generalise to unseen data was inadequate relative to MLP and PointNet.
Whilst these results for the latter models were relatively good, MLP showed far more accuracy and consistency across the separate experiments, with the total RMSE found at surface, fluid, and $C_L$ decreasing with every increase of aerofoil number. To fully investigate the extent of this improving accuracy, research should be undertaken to verify the difference training on more than 150 aerofoils would make.

In consideration of future work, one possibility as to why the models are over-fitting is that they are being overtrained with too many epochs. While increasing the number of epochs will reliably decrease the training and testing losses, further study is required in order to provide further insight into its effect on the validation process.



\bibliographystyle{IEEEtran}
\balance
\bibliography{sample}

\end{document}